\documentclass[12pt,prd,onecolumn,showpacs,amsmath,amssymb,aps,floats,floatfix,nofootinbib]{revtex4-1}
\usepackage[colorlinks=true,urlcolor=blue,anchorcolor=blue,citecolor=blue,filecolor=blue,linkcolor=blue,menucolor=blue,pagecolor=blue,linktocpage=true]{hyperref} % should be commented out if the tex file will be compiled with latex in arXiv (pdflatex is fine)

\usepackage[inline]{enumitem}
\usepackage[multidot]{grffile} % allow the name of figures to include dots
\usepackage{dcolumn}
\usepackage{bm}
\usepackage{amsmath}
\usepackage{amsfonts}
\usepackage{amssymb}
\usepackage{color}
\usepackage{latexsym}
\usepackage{slashed} % slash mark
\usepackage{pstricks}
\usepackage{indentfirst}
\usepackage{mathrsfs}
\usepackage{multirow}
\usepackage{epsfig,psfrag}
\usepackage{subfigure}
\usepackage{mathtools}
\usepackage{setspace} % spacing
\usepackage[utf8]{inputenc} % accept utf-8 input encoding
\usepackage[scientific-notation=true]{siunitx} % comprehensive units
\usepackage{verbatim}

\graphicspath{{fig/}}

\makeatother

\allowdisplaybreaks % allow eqnarray breaks
\begin{document}
\title{Freeze-in Production of Pseudo-Nambu-Goldstone Dark Matter Model with a Real Scalar }
\author{Xue-Min Jiang}
\author{Chengfeng Cai}\email{caichf3@mail.sysu.edu.cn}
\author{Yu-Hang Su}
\author{Hong-Hao Zhang}\email{zhh98@mail.sysu.edu.cn}
\affiliation{School of Physics, Sun Yat-Sen University, Guangzhou 510275, China}

\begin{abstract}
In this work, we study a pseudo-Nambu-Goldstone boson (pNGB) dark matter model extended with a real scalar.
The dark sector is assumed to be feebly coupled with the standard model (SM) via a Higgs portal, so that the pNGB dark matter is produced by the freeze-in mechanism. Since the production happened in a very high energy era, we introduce an extra scalar field which is weakly coupled to the SM for stablizing the electroweak vacuum. Our model can reproduce the correct relic abundance of dark matter favored by observations.
In addition, we determine the evolution of couplings in higher energy scale by solving the renormalization group equations, and show that the self coupling of the real scalar, $\lambda_S$, and the mixing coupling between the Higgs field and the real scalar, $\lambda_{HS}$ are stringently constrained by the conditions of vacuum stability and couplings perturbativity up to the Planck scale. We also find that the relic abundance of DM is insensitive to the values of $\lambda_S$ and $\lambda_{HS}$ unless the dominant production processes are $\overline{H}+H(S+S)\to\phi+\phi$ via t- and u-channels.

\end{abstract}

\maketitle
%\tableofcontents
%\clearpage

\section{Introduction}
A lot of evidences from cosmology and astrophysics have shown that a large fraction of energy in our universe is consists of dark matter.
However, the nature of dark matter (DM) still remains a mystery.
One of the most studied candidate of DM is the weakly interacting massive particle (WIMP), which can naturally approach the observed relic abundance via the thermal freeze-out mechanism~\cite{Gondolo:1990dk}.
In the freeze-out scenario, DM particles weakly couple to the standard model (SM) particles, and thus they are in thermal equilibrium with the plasma during the radiation dominant era. As the temperature drops down, the annihilation rate of DM becomes too small to defeat the Hubble expansion, then DM particles stop annihilating with each other and their number density in the comoving frame tends to a fixed value. In the recent decades, there were many experiments tried to detect WIMPs directly~\cite{LUX:2016ggv,XENON:2018voc,PandaX-4T:2021bab,LUX-ZEPLIN:2022qhg} and indirectly~\cite{MAGIC:2016xys,CTAConsortium:2012fwj}, however no persuasive signal was obtained yet, and thus the scenario of WIMPs freeze-out production is facing more challenges. In order to naturally explain the observed relic relic abundance of DM without violating the direct detection constraints, people have proposed lots of strategies. An appealing model is the pseudo-Nambu-Goldstone boson (pNGB) DM which is produced by the traditional freeze-out mechanism~\cite{Gross17,Jiang:2019soj, Liu:2022evb, Cai:2021evx, Zhang:2021alu, Arina:2019tib,Abe:2020iph, Okada:2021qmi,Glaus:2020ihj,Okada:2021qmi,Darvishi:2022wnd,Abe:2022mlc}. In this case, direct detection signal can be naturally suppressed due to an automatic cancellation among the amplitude for DM-nucleon scattering in the limit of zero-momentum transfer.

Recently, pNGB DM model is also studied under the freeze-in production scenario~\cite{Abe20,Sakurai:2021ipp,Kondo:2022lgg}, which is an alternative paradigm that is capable to evade the stringent direct detection bound. The freeze-in production mechanism requires the DM candidate to feebly interact with the SM particles, so that DM never get into thermal equilibrium with the SM plasma. Particles with such a feeble interaction property is usually called feebly interacting massive particles (FIMPs)~\cite{Hall10,Bernal:2017kxu}. The freeze-in mechanism assumes that DM candidates have negligible density after the reheating, and then they are produced via annihilation or decay of the SM plasma. Due to the feebleness of interactions, DM candidate can easily circumvent current direct and indirect detection constraints. In recent years, searching FIMPs have gathered more and more attention~\cite{Hambye:2018dpi,Belanger:2018sti,Brooijmans:2020yij,Calibbi:2021fld,Dvorkin:2020xga,No:2019gvl,Ghosh:2022fws,Elor:2021swj,Bhattiprolu:2022sdd}.

In the framework of pNGB DM, feeble couplings can be easily achieved by assuming a large vacuum expectation value (VEV) of a complex scalar field, $\Phi$. The phase component of $\Phi$ is a Nambu-Goldstone boson if there is a global U(1) symmetry in the model. The phase component of $\Phi$ can obtain an arbitrary size of mass if the U(1) symmetry is softly broken, then it becomes a pNGB and its mass can be naturally small comparing to the VEV of $\Phi$ and the electroweak scale. Note that the freeze-in production processes usually happen in a very high energy era, it reminds us to worry about the stability of the electroweak vacuum~\cite{Bezrukov:2012sa, Buttazzo:2013uya, Degrassi:2012ry}. After the reheating of the universe, there are thermal corrections to the Higgs potential which can fill the dip of the effective potential, so the vacuum of Higgs field is safe in this stage. In an earlier stage during a high scale inflation, Higgs can fluctuate in an order of the Hubble parameter, therefore, it is probable to decay into the lower energy vacuum~\cite{Espinosa:2007qp,Kobakhidze:2013tn,Fairbairn:2014zia,Hook:2014uia,Kamada:2014ufa,Herranen:2014cua,Kearney:2015vba,Espinosa:2015qea}. One way to solve this problem is to identify the $\Phi$ field of the pNGB model as the inflaton field, and then the Higgs field receives an effective mass due to a Higgs portal interaction. If this effective mass is large enough during the inflation, the fluctuation of Higgs field can be suppressed. However, during the preheating era after inflation, a large fluctuation is still possible to be generated due to a broad resonance of Higgs~\cite{Ema:2016kpf}. According to the study in Ref.\cite{Ema:2016kpf}, the coupling between the Higgs field and the inflaton field has a stringent upper bound if the metastablity of Higgs vacuum need to survive.

Motivated by this subtlety, we are going to consider another strategy of stabilizing electroweak vacuum. It is well known that the instability of the electroweak vacuum is caused by the fact that the Higgs coupling, $\lambda_H(\mu)$, runs to a negative value in the high energy scale. A simple solution to this problem is introducing some new physics which couple with the Higgs field below the scale of $\lambda_H$ becoming negative~\cite{Falkowski:2015iwa, Chen:2014ask, Elias-Miro:2012eoi,Gonderinger:2012rd,Gabrielli:2013hma,Khoze:2014xha,Ghorbani:2021rgs}. In this work, we will consider the simplest model which only extend the pNGB model with a real scalar field, denoted as $S$. The scalar field $S$ can stabilize the vacuum by two effects. One effect is that the coupling between the Higgs field and $S$, denoted as $\lambda_{HS}$, can modify the beta-function of $\lambda_H$ and slow down the dropping of $\lambda_H$ in the high energy~\cite{Silveira:1985rk,Chen:2012faa}. The other one is the threshold effect of $\lambda_H$ around the mass scale of introducing $S$ due to the mixing between the Higgs and $S$~\cite{Randjbar-Daemi:2006ada,Elias-Miro:2012eoi}. Usually, a stable electroweak vacuum requires a sizable $\lambda_{HS}(m_S)$ at the threshold scale $m_S$. However, there is another theoretical constraint for the sizes of couplings, which is the requirement of perturbativity below the cutoff scale. If the cutoff scale is to chosen as the Planck scale, it usually put a stringent upper bound on the quartic couplings of the potential terms, and thus the available parameter space will be quite restricted. It is also natural to assume a feeble interaction between $S$ and $\Phi$, then the dark sector is possible to be produced via the annihilation of a pair of $S$ particles in the framework of freeze-in mechanism as well. In this work, we will focus on the IR freeze-in scenario, and then discuss how does the scalar $S$ affect the freeze-in production of DM.

The paper is organized as follow.
In section~\ref{se.pNGB+S}, we introduce the pNGB+$S$ dark matter model.
In section~\ref{se.running}, the renormalization group equations (RGEs) for couplings are solved numerically, and the constraints from the perturbativity of couplings and vacuum stability are studied.
The freeze-in production of pNGB DM is discussed in section~\ref{se.freezein}, and finally we give a summary in section~\ref{se.summary}. In appendix~\ref{app1}, we compare the results with and without the t- and u-channels of $\overline{H}+H\to\phi+\phi$ process. In appendix~\ref{se.beta}, we show the beta functions of couplings in the SM.

\section{A pNGB+S Dark Matter Model } \label{se.pNGB+S}

The basic setup of our model is to extend the SM with a complex scalar, $\Phi$, which supplies a pNGB as the DM candidate~\cite{Gross17}, and a real scalar, $S$, which helps the vacuum becoming stable.
We further assume a $Z_2$ symmetry under which $\Phi\to-\Phi,~S\to-S$ for simplicity, then the Lagrangian for the scalar fields reads
\begin{eqnarray}
\mathcal{L}&=&\frac{1}{2}\partial_\mu S \partial^\mu S +(\partial_\mu \Phi)^\dagger\partial^\mu \Phi +D_\mu H^\dagger D^\mu H -V_0(H,\Phi,S),\\
V_0(H,\Phi,S)&=&-\frac{1}{2}\mu_S^2 S^2 +\frac{1}{4}\lambda_S S^4-\mu_0^2 |H|^2+\lambda_H|H|^4\nonumber \\
&&\quad -\mu_\Phi^2|\Phi|^2+\lambda_\Phi|\Phi|^4+\frac{1}{2}\lambda_{HS} |H|^2 S^2+\frac{1}{2}\lambda_{\Phi S}|\Phi|^2 S^2+\lambda_{H \Phi} |H|^2 | \Phi |^2\nonumber \\
&& \quad-\frac{1}{4}\mu_\Phi^{\prime 2}[\Phi^2 + (\Phi^*)^2],\label{potential0}
\end{eqnarray}
where $H$ is the SM Higgs field, and $V_0(H,\Phi,S)$ is the tree-level zero temperature potential terms.
Note that the last term in $V_0(H,\Phi,S)$ softly breaks a global $U(1)$ symmetry for $\Phi$, and then generates a mass for the the phase component of $\Phi$, which is a pNGB.
Due to the chosen sign of the mass terms in Eq.\eqref{potential0}, $H$, $\Phi$, and $S$ will develop non-vanishing VEVs  $\langle H\rangle=(0,v/\sqrt{2})^T$, $\langle\Phi\rangle =v_\phi/\sqrt{2}$ and $\langle S\rangle=w$, respectively. In zero temperature, these VEVs spontaneously break the gauged $\mathrm{SU}(2)_L\times\mathrm{U}(1)_Y$, and the global $U(1)\times Z_2$ symmetries (we dub it as the broken phase). Using the unitary gauge, we can parameterize $H$, $\Phi$, and $S$ as follows,
\begin{eqnarray}
H(x) = \frac{1}{\sqrt{2}} \begin{pmatrix}0\\ h(x) + v\end{pmatrix},\quad S(x) =w + s(x),\quad \Phi(x) =\frac{v_\phi+\phi(x)}{\sqrt{2}}e^{i\chi(x)/v_\phi}.
\end{eqnarray}
The VEVs of the scalars are determined by the stationary point conditions:
\begin{eqnarray}\label{spc}
\mu_0^2  &=& \lambda_Hv^2+ \frac{1}{2}\lambda_{HS}  w^2+\frac{1}{2}\lambda_{H \Phi}   v_\phi^2~, \label{mu02}\\
\mu_S^2 &=& \lambda_S w^2+\frac{1}{2}\lambda_{HS} v^2 +\frac{1}{2}\lambda_{\Phi S}v_\phi^2~,\label{muS2} \\
\mu_\Phi^2&=&\lambda_\Phi v_\phi^2+\frac{1}{2}\lambda_{\Phi S}w^2+\frac{1}{2}\lambda_{H \Phi}v^2 -\frac{1}{2}\mu_\Phi^{\prime 2}~.
\end{eqnarray}
After the spontaneous symmetry breaking, a remnant $Z_2$ symmetry ($\Phi\to\Phi^\ast$) at tree-level can ensure the pNGB DM candidate to be cosmologically stable.
The mixing mass matrix for $(h,s,\phi)$ is given by
\begin{eqnarray}
\mathcal{M}^2=\begin{pmatrix}2\lambda_Hv^2&\lambda_{HS} v w&\lambda_{H \Phi}  v v_\phi\\ \lambda_{HS} v w&2\lambda_S w^2&\lambda_{\Phi S}v_\phi w\\ \lambda_{H \Phi}  v v_\phi&\lambda_{\Phi S}v_\phi w&2\lambda_\Phi v_\phi^2\end{pmatrix}\label{MS2}~.
\end{eqnarray}
In this work, we assume that $\Phi$ feebly couples to both $H$ and $S$, with couplings $\lambda_{H\Phi},\lambda_\Phi,\lambda_{\Phi S}\ll1$ which satisfy $(\lambda_{H\Phi}^2/\lambda_\Phi)\ll\lambda_H,\lambda_S$, then the mass-squared matrix can be decomposed into a 2 by 2 matrix in $(h,s)$ basis as:
\begin{eqnarray}
\mathcal{M}_{2}^2=\begin{pmatrix}2\lambda_Hv^2&\lambda_{HS} v w\\ \lambda_{HS} v w&2\lambda_S w^2\end{pmatrix}\label{MS3}~.
\end{eqnarray}
and a mass-squared for $\phi$ as
\begin{eqnarray}
m_{\phi}^2\approx 2\lambda_\Phi v_\phi^2.
\end{eqnarray}
Eq.\eqref{MS3} can be diagonalized by a two by two orthogonal matrix $O$ as
\begin{eqnarray}
\mathcal{M}_{2}^2\to O\mathcal{M}_{2}^2O^T=\mathrm{diag}\{m_1^2,m_2^2\}~,
\end{eqnarray}
where
\begin{eqnarray}
m_{1,2}^2=\lambda_Hv^2+\lambda_Sw^2\mp\sqrt{(\lambda_Hv^2-\lambda_Sw^2)^2+\lambda_{HS}^2v^2w^2}.
\end{eqnarray}

In the high temperature, the 1-loop finite temperature corrections to the potential has a significant effect. It effectively induces corrections to the masses of the Higgs field and the real scalar $S$. These corrections can compensate the negative mass-squared parameters and thus lead to the restoration of electroweak and $Z_2$ symmetries.
The leading 1-loop finite temperature corrections to the potential is given by~\cite{Qurios99}
\begin{eqnarray}
V_{th}\approx\frac{T^4}{2\pi^2}\left[\sum_{i=Z,W^\pm,h^0, h^\pm ,s}n_i J_B(\hat m_i^2/T^2)-n_t J_F(\hat m_t^2/T^2)
\right],
\end{eqnarray}
where $n_i$ is the degree of freedom (d.o.f.) of the corresponding bosonic particle and $n_t=12$ is the d.o.f. of top quark. $J_B$ and $J_F$ are the thermal bosonic and fermionic function defined by~\cite{Qurios99},
\begin{eqnarray}
J_{B,F}\left[m_i^2/T^2\right]=\int_0^{\infty} d x x^2 \ln \left[1\mp e^{-\sqrt{x^2+m_i^2/T^2}}\right]~.
\end{eqnarray}
Note that $\hat m_{i}^2(h^0, s)$ and $\hat m_{t}^2(h^0, s)$ are the tree-level background fields dependent mass-squared for $i$-th boson and top quark which will be derived later.
In high temperature limit, the potential can be expanded as~
\begin{eqnarray} \label{Vth}
V_{th}&=&\frac{T^4}{2\pi^2}\left[\sum_{i= Z,W^\pm,h^0,h^\pm,s}n_i \left(-\frac{\pi^4}{45} + \frac{\pi^2}{12}\frac{\hat m_i^2}{T^2}\right)+n_t \left(-\frac{7 \pi^4}{360} + \frac{\pi^2}{24}\frac{\hat m_t^2}{T^2}\right)\right]+\mathcal{O}(T)~.
\end{eqnarray}
The $T^2$ terms in Eq.\eqref{Vth} are the leading finite temperature corrections to the background fields masses, and thus they can stabilize the EW symmetry restoring vacuum at the origin of Higgs fields when the corrections have a positive sign.

Now we shall derive the tree-level mass eigenvalue for each d.o.f.. We should parametrize the Higgs doublet and $S$ as $H=(h^\pm,(h^0+i\eta)/\sqrt{2}), S=s$, and treat $h^0$ and $s$ as the background fields.
From the tree-level potential Eq.\eqref{potential0}, $s$ mixes with $h^0$.
The background field dependent mass matrix in the $(h^0,s)$ basis is defined as
\begin{eqnarray}
V_0&\supset&\frac{1}{2}\begin{pmatrix}h^0 & s\end{pmatrix}
\begin{pmatrix}
\hat M^2_{11}& \hat M^2_{12}\\ \hat M^2_{21}&\hat M^2_{22}
\end{pmatrix}
\begin{pmatrix}
h^0 \\s
\end{pmatrix},
\end{eqnarray}
where
\begin{eqnarray}
\hat{M}^2_{11}&=&\frac{\partial^2 V_0}{\partial (h^0)^2}=\frac{1}{2}\lambda_{HS}  s^2+\lambda_{H \Phi} \frac{v_\phi^2}{2}-\mu_0^2+3 \lambda_H( h^0 )^2~,\nonumber \\
\hat{M}^2_{12}&=&\hat M^2_{21}=\frac{\partial^2 V_0}{\partial s \partial h^0}=\lambda_{HS}  h^0 s~, \\
\hat{M}^2_{22}&=&\frac{\partial^2 V_0}{\partial s^2}=-\mu_S^2 +3\lambda_S s^2+\frac{1}{2}\lambda_{HS} (h^0)^2 +\frac{1}{2}\lambda_{\Phi S}v_\phi^2~.\nonumber
\end{eqnarray}
The masses of $h^\pm$ and $\eta$ can be directly read off as
\begin{eqnarray}\label{mhetahat}
\hat{m}^2_{h^\pm}&=&\frac{1}{2}\lambda_{HS}  s^2+ \frac{1}{2}\lambda_{H \Phi}v_\phi^2+\lambda_H(h^0)^2-\mu_0^2,\nonumber \\
\hat m^2_{\eta}&=&\frac{1}{2}\lambda_{HS}  s^2+ \frac{1}{2}\lambda_{H \Phi}v_\phi^2+\lambda_H(h^0)^2-\mu_0^2.
\end{eqnarray}
The gauge fields and top quarks also receive masses due to the background fields as follows,
\begin{eqnarray}
m_W^2=\frac{1}{4}g^2(h^0)^2,~m_{Z}^2=\frac{1}{4}(g^2+g'^2)(h^0)^2,~m_t^2=\frac{1}{2}y_t^2(h^0)^2.
\end{eqnarray}
The finite temperature corrections to $H$ and $S$ in high temperature can be easily obtained by using the expansion in Eq.\eqref{Vth}:
\begin{eqnarray}	
\delta_Tm_H^2&\approx&\frac{T^2}{4}\left(\frac{1}{4}g^{\prime 2}+\frac{ 3}{4}g^2+y_t^2+\frac{ 1}{6}\lambda_{HS}+2\lambda_{H}\right)~,\\
\delta_T m_S^2&\approx&\frac{ T^2}{4}
\left(\frac{2}{3}\lambda_{HS}+ \lambda_S\right),
\end{eqnarray}
where $g^\prime$, and $g$ are $U(1)_{Y}$, $SU(2)_L$ gauge couplings, while $y_t$ is the top Yukawa coupling.

In the era of reheating, the temperature is so high that the EW symmetry and the $Z_2$ of $S$ are restored due to the finite temperature corrections of the potential, while the global $U(1)$ remains broken since $\Phi$ is feebly couple to the thermal bath. The masses for the components in $\Phi=(v_\phi+\phi)e^{i\sqrt{2}\chi/v_\phi}/\sqrt{2}$ are simply given by the zero temperature ones as
\begin{eqnarray}
m_\phi^2=\lambda_\Phi v_\phi^2,\quad m_\chi^2=\mu^{\prime 2}_\Phi.
\end{eqnarray}

\section{Constraints from perturbativity and vacuum stability}\label{se.running}
We will make use of the perturbativity of couplings and the vacuum stability to constrain our model. The tools for our analysis is the renormalization group equations (RGE) of the couplings. The condition of perturbative couplings requires all the couplings keeping small than $\sim4\pi$ up to the Planck scale, while the vacuum stability requires some combination of couplings in the potential to keep positive up to the Planck scale. These conditions will stringently constrain the values of couplings in the low energy scale. Since the freeze-in production happens after the reheating era, our calculation of the yield of DM should include the effects of running couplings.

In order to study the evolutions of the couplings, we consider the $\beta$-functions of the SM+$S$ model.
The most relevant $\beta$-functions are the gauge and Yukawa couplings in the SM, and the quartic couplings in the potential such as $\lambda_H$, $\lambda_S$, and $\lambda_{HS}$. The 1-loop $\beta$-functions for these quartic couplings are given by~\cite{Gonderinger:2009jp}
	\begin{eqnarray}
	\beta_{\lambda_{H}} &=&\beta_{\lambda_H}^{\textrm{SM}}+\frac{ \lambda_{HS}^2}{32 \pi^2} ,\label{betaH}\\
	\beta_{\lambda_{S}} &=&	\frac{1}{16 \pi^{2}}\left[2 \lambda_{ HS}^{2}+18 \lambda_{S}^{2}\right] ,\label{betaS}\\
	\beta_{\lambda_{HS}} &=&\frac{1}{16 \pi^{2}}\left[4 \lambda_{HS}^{2}+12 \lambda_H \lambda_{HS}+6 \lambda_{S}\lambda_{HS}+ 6\lambda_{HS}y_t^2-\lambda_{HS}\left(\frac{9}{2} g^{2}+\frac{3}{2} g^{\prime 2}\right)\right].\label{betaHS}
	\end{eqnarray}
The other $\beta$-functions for the SM are summarized in appendix~\ref{se.beta}. Note that in principle, we also need to consider the running of $\lambda_\Phi$, $\lambda_{H\Phi}$, and $\lambda_{S\Phi}$ from the scale of $v_\phi$ to the Planck scale. However, their runnings are negligible when we focus on the case of freeze-in production of DM, since these couplings have very tiny values.

According to Eq.\eqref{betaS}, $\lambda_S$ always grow as the energy increase, and then it will finally blow up if the theory is not cutoff. Even if $\lambda_S$ is set to be vanishing at some low energy scale, it can be generated by a non-vanishing $\lambda_{HS}$ when it runs to higher energy scale. Therefore, we can obtain a stringent upper bound for $\lambda_S$ and $\lambda_{HS}$ by requiring them to keep perturbative below the Planck scale. To be precise, the constraints we impose on the couplings are
\begin{eqnarray}
g_i(\mu),y_t(\mu),\lambda_S(\mu),\lambda_H(\mu),|\lambda_{HS}(\mu)|<4\pi,\label{pertcond}
\end{eqnarray}
for $m_S\sim \textrm{TeV}<\mu<M_P$.

%%%%%%%%%%%%%%%%%%%%%%%%%%%%%%%%%%%%%%%%%%%%%%%%%
Another problem raised by the running couplings is the unstable vacuum caused by the SM Higgs field. It is well known that the quartic coupling $\lambda_H$ of the SM Higgs field evolves to a negative value as the energy scale reaches $\sim 10^{10}$~GeV. This leads to an unstable (or metastable) vacuum since the potential has a lower minimum with a larger VEV of the Higgs field.
In our model, the existence of a new scalar singlet $S$ can help to stabilize the vacuum~\cite{Falkowski:2015iwa, Gonderinger:2009jp,Elias-Miro:2012eoi,Gonderinger:2012rd,Gabrielli:2013hma,Khoze:2014xha,Chen:2014ask,Ferreira:2004yd,Ema:2017ckf,Salvio:2015cja,Salvio:2018rv}. One effect comes from the loop contribution of the singlet scalar to the $\beta$-function of the Higgs quartic coupling $\lambda_H$~\cite{Chen:2012faa,Gonderinger:2012rd,Gonderinger:2009jp}, as we can see from the second term in the right hand side of Eq.\eqref{betaH}.
Another important correction comes from the threshold effect originated from the mixing between the Higgs doublet and the real singlet~\cite{Randjbar-Daemi:2006ada,Elias-Miro:2012eoi}.  Considering a situation that the threshold scale of the real singlet, $m_S\sim \sqrt{\lambda_S} w$, is much larger than the electroweak scale, $m_h\sim \sqrt{\lambda_H} v$, then one can integrate out the field $S$ below a scale about $m_S$. This leads to an effective quartic operator in the zero temperature potential as $V_{\mathrm{eff}}\supset\tilde{\lambda}_H(\mu) |H|^4$ with a matching condition at $m_S$:
\begin{eqnarray}
\tilde{\lambda}_H(m_S)=\lambda_{H}(m_S)-\frac{1}{4}\frac{\lambda_{H S}^{2}(m_S)}{\lambda_{S}(m_S)}~.\label{matching}
\end{eqnarray}
%%%%%%%%%%%%%%
In practice, we start with the pure SM couplings at the electroweak scale $m_Z$, and use the pure SM $\beta$-functions to determine the evolutions of those couplings cutoff at the threshold scale $m_S$. At the scale $\mu=m_S$, we use the matching condition Eq.\eqref{matching} to obtain the UV Higgs quartic coupling $\lambda_H$ related to the effective Higgs quartic coupling $\tilde{\lambda}_H$. Finally, we determine the evolutions of couplings by solving the $\beta$-functions listed in appendix~\ref{se.beta} together with Eqs.\eqref{betaH}-\eqref{betaHS} in the range $m_S<\mu<M_P$. On the other hand, the vacuum stability requires the quartic couplings to satisfy
\begin{eqnarray}
 \lambda_H(\mu)>0,\quad  \lambda_S(\mu)>0\quad \lambda_{ HS}(\mu)>-2\sqrt{ \lambda_H(\mu)\lambda_S(\mu)}~.\label{VScond}
\end{eqnarray}
in the range of $m_S<\mu<M_P$.

\begin{figure}[!t]
	\centering
	\subfigure[\label{fig5-1}]
	{\includegraphics[width=0.32\textwidth]{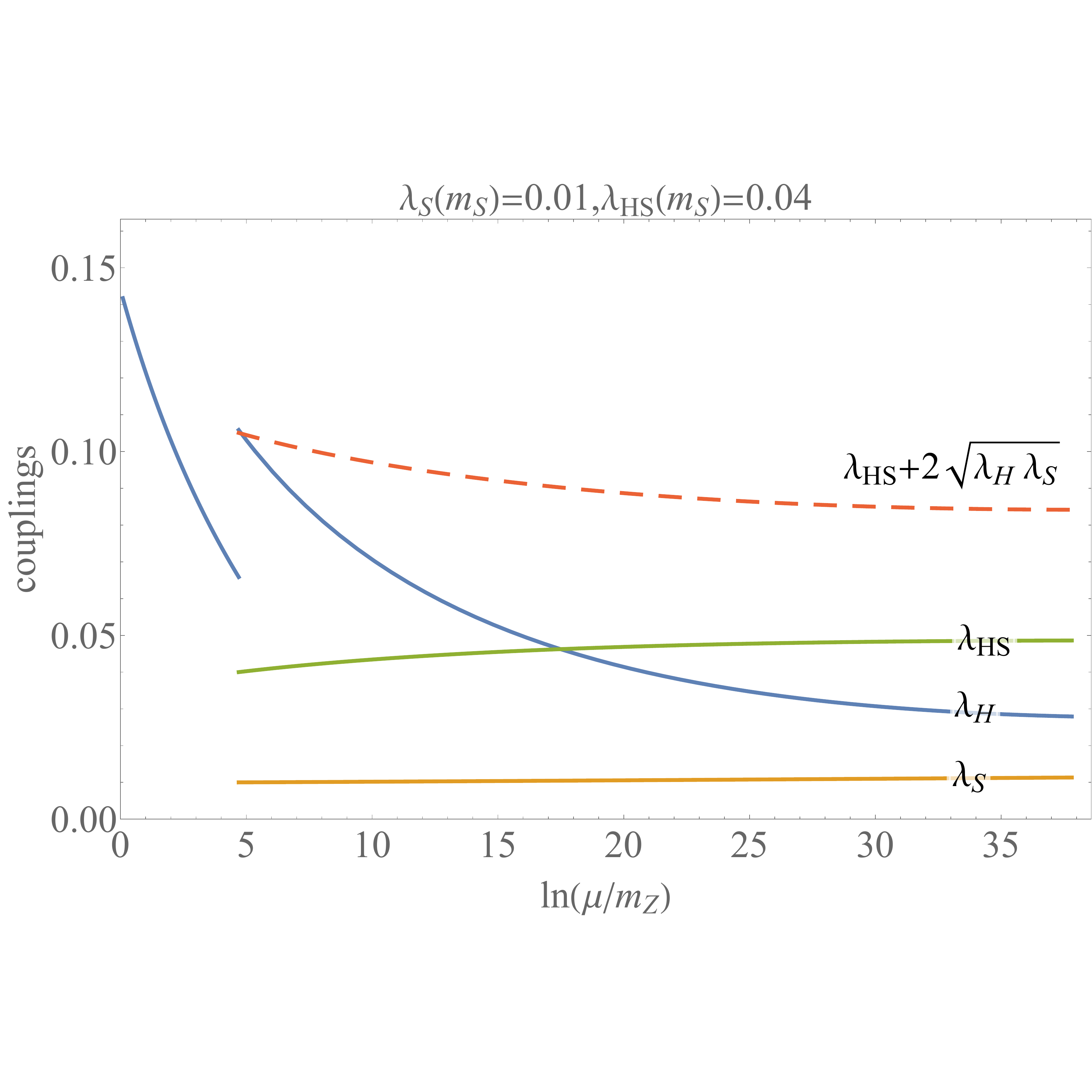}}
	\subfigure[\label{fig5-2}]
	{\includegraphics[width=0.32\textwidth]{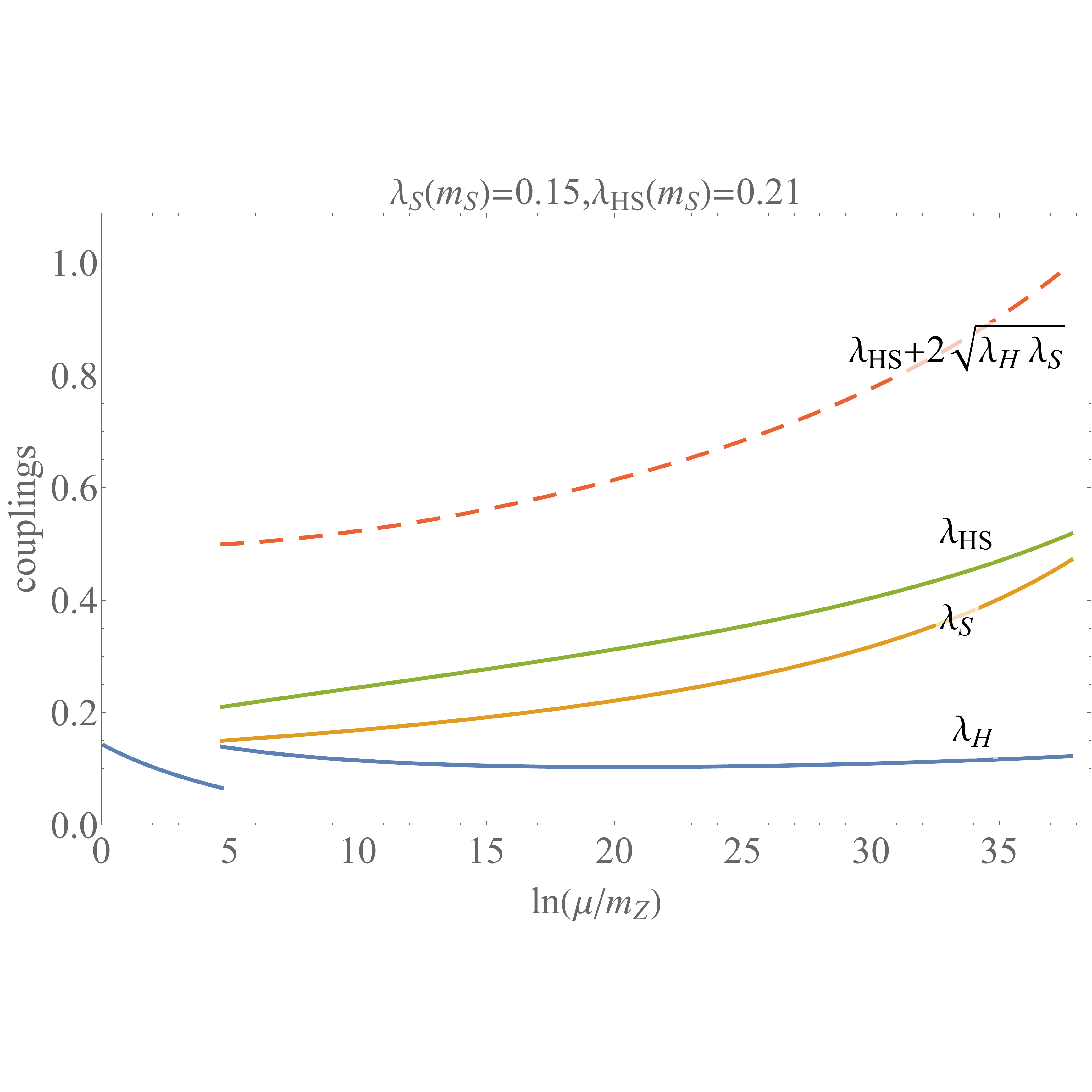}}
	\subfigure[\label{fig5-3}]
	{\includegraphics[width=0.32\textwidth]{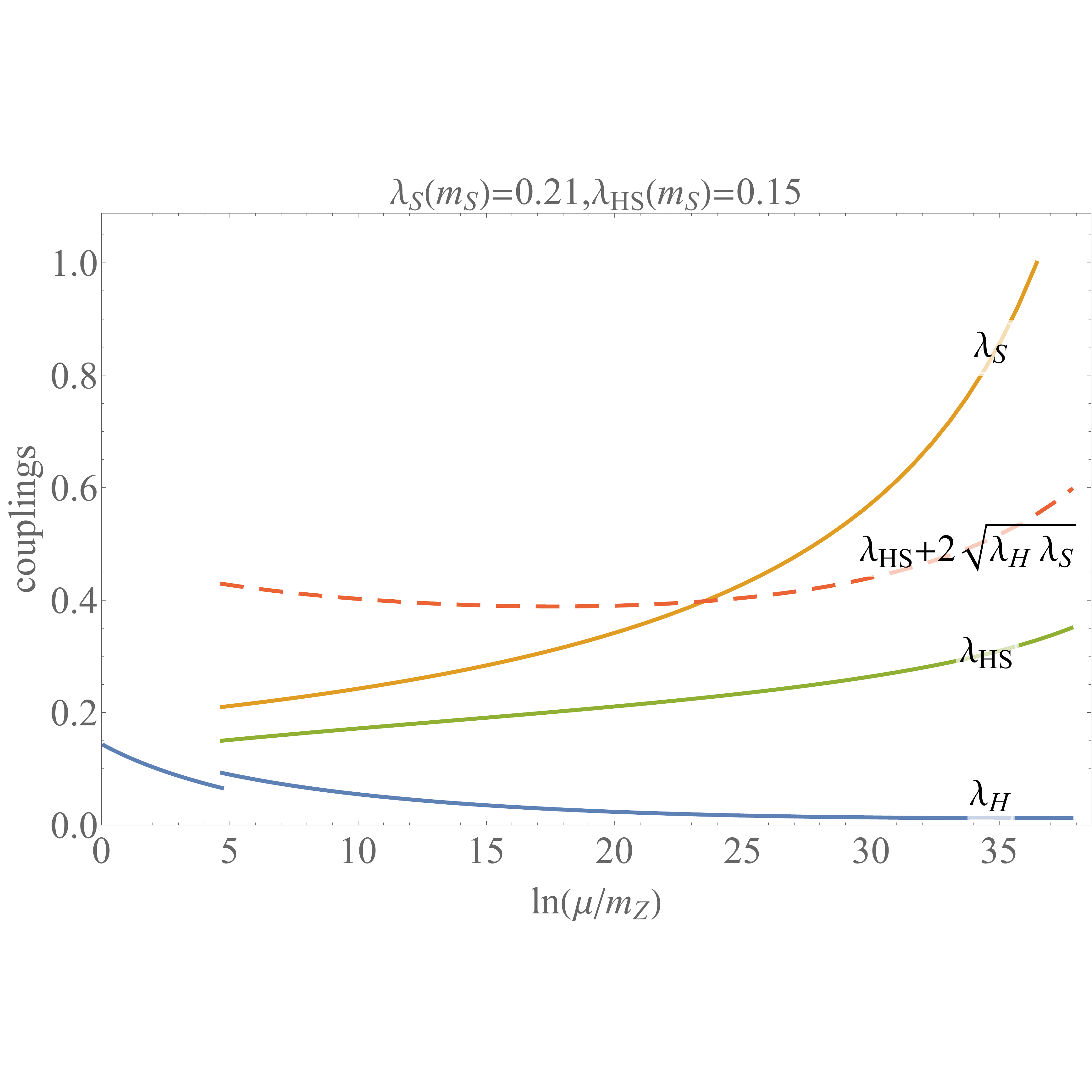}}
	\subfigure[\label{fig5-4}]
	{\includegraphics[width=0.32\textwidth]{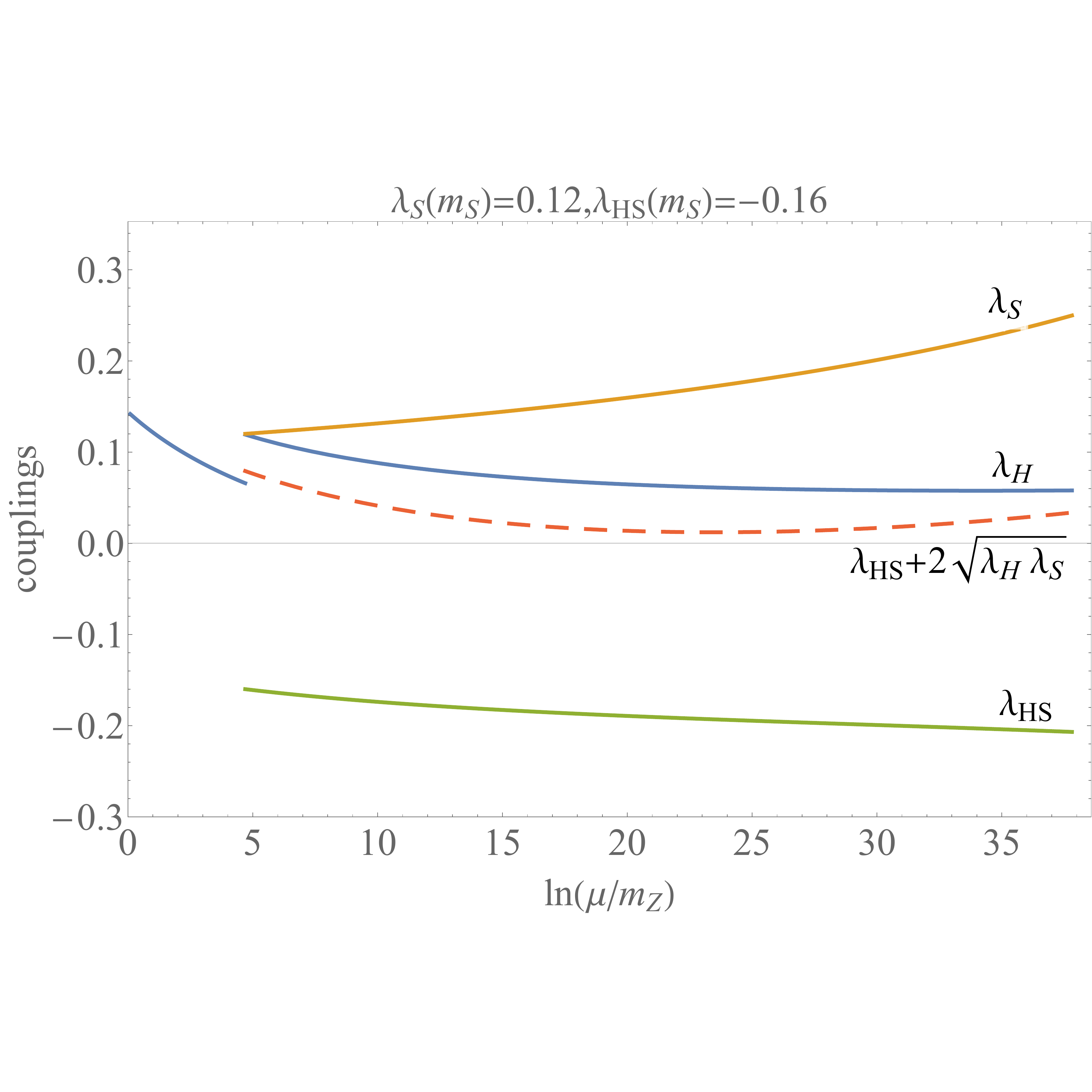}}
	\subfigure[\label{fig5-5}]
	{\includegraphics[width=0.32\textwidth]{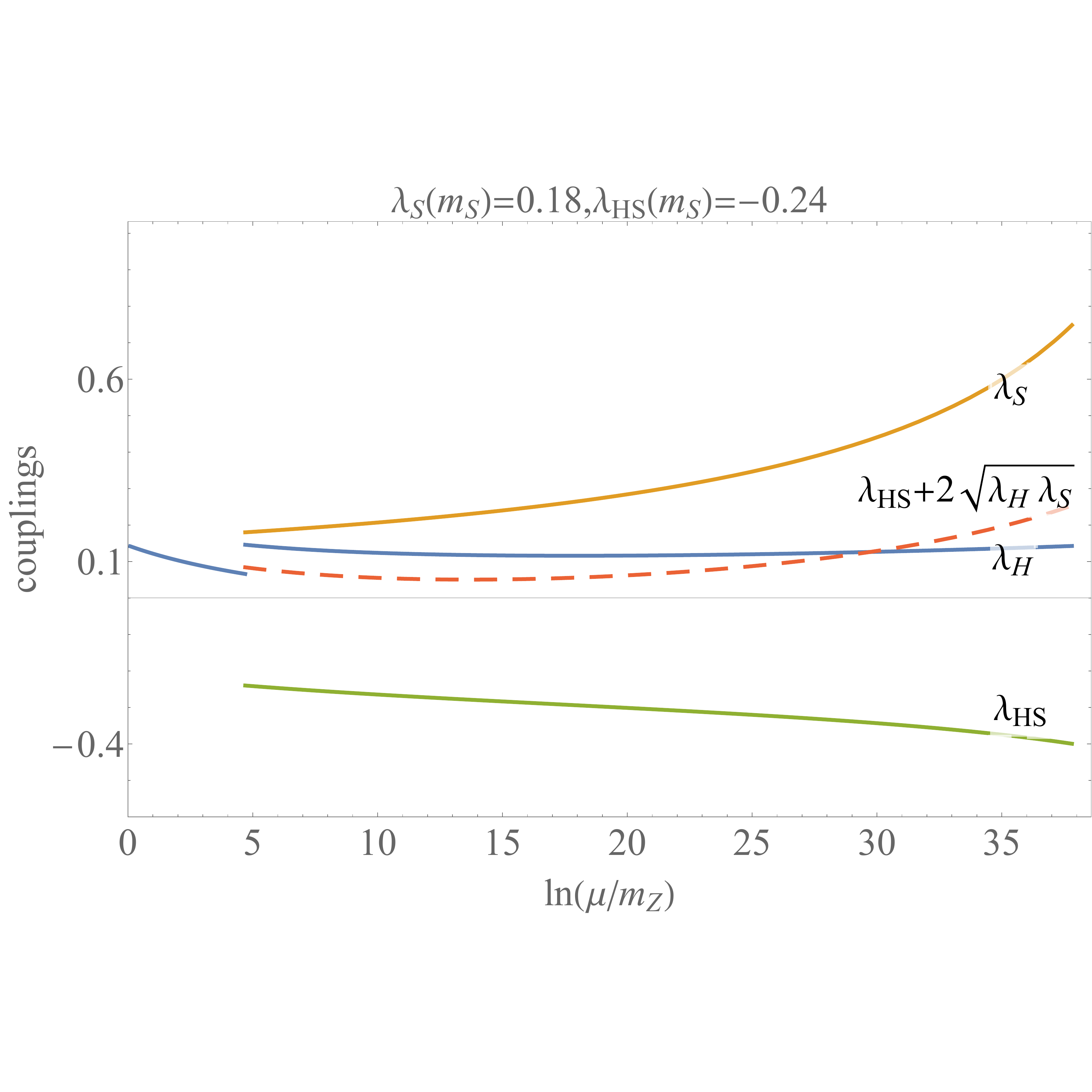}}
	\subfigure[\label{fig5-6}]
	{\includegraphics[width=0.32\textwidth]{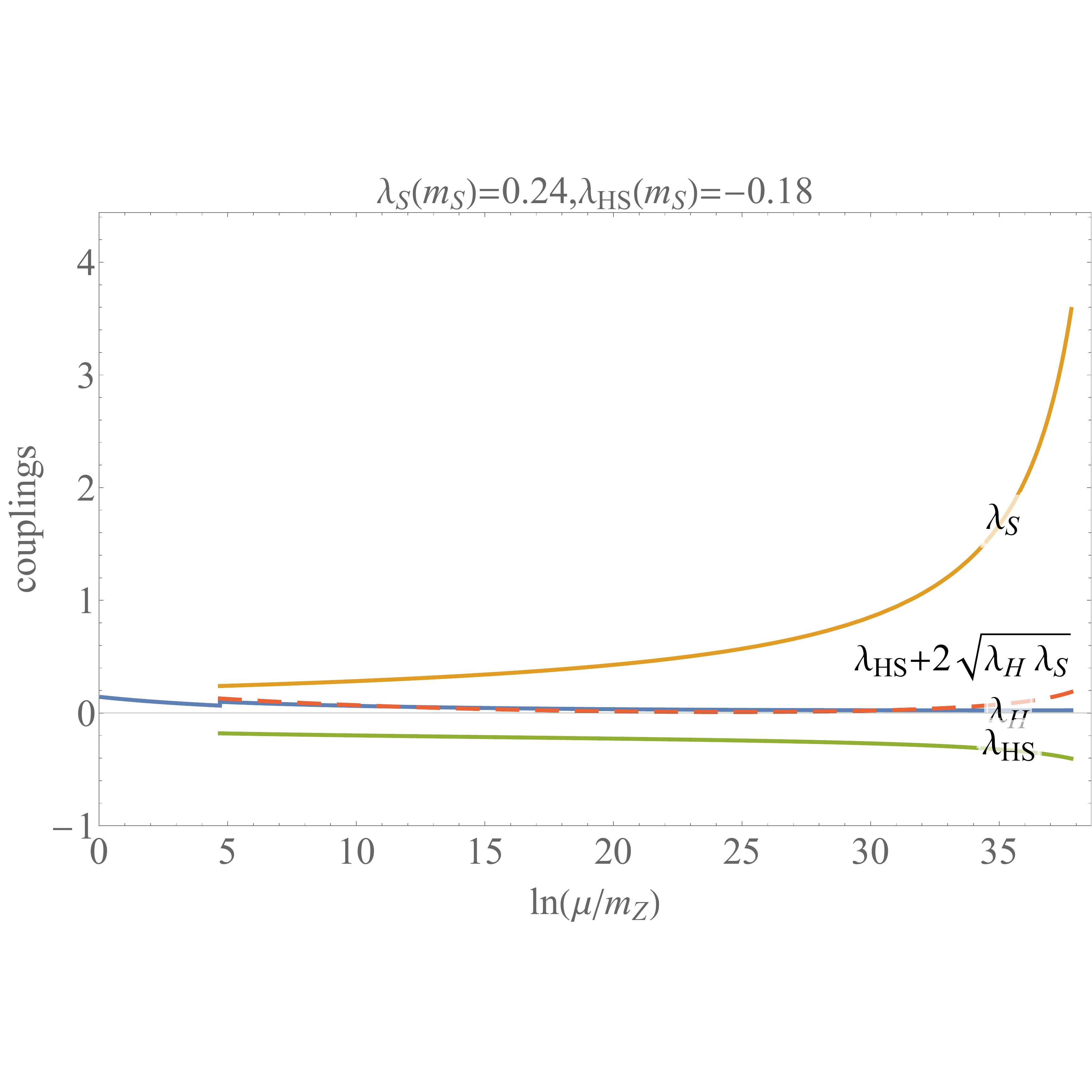}}
		\subfigure[\label{fig5-7}]
	{\includegraphics[width=0.32\textwidth]{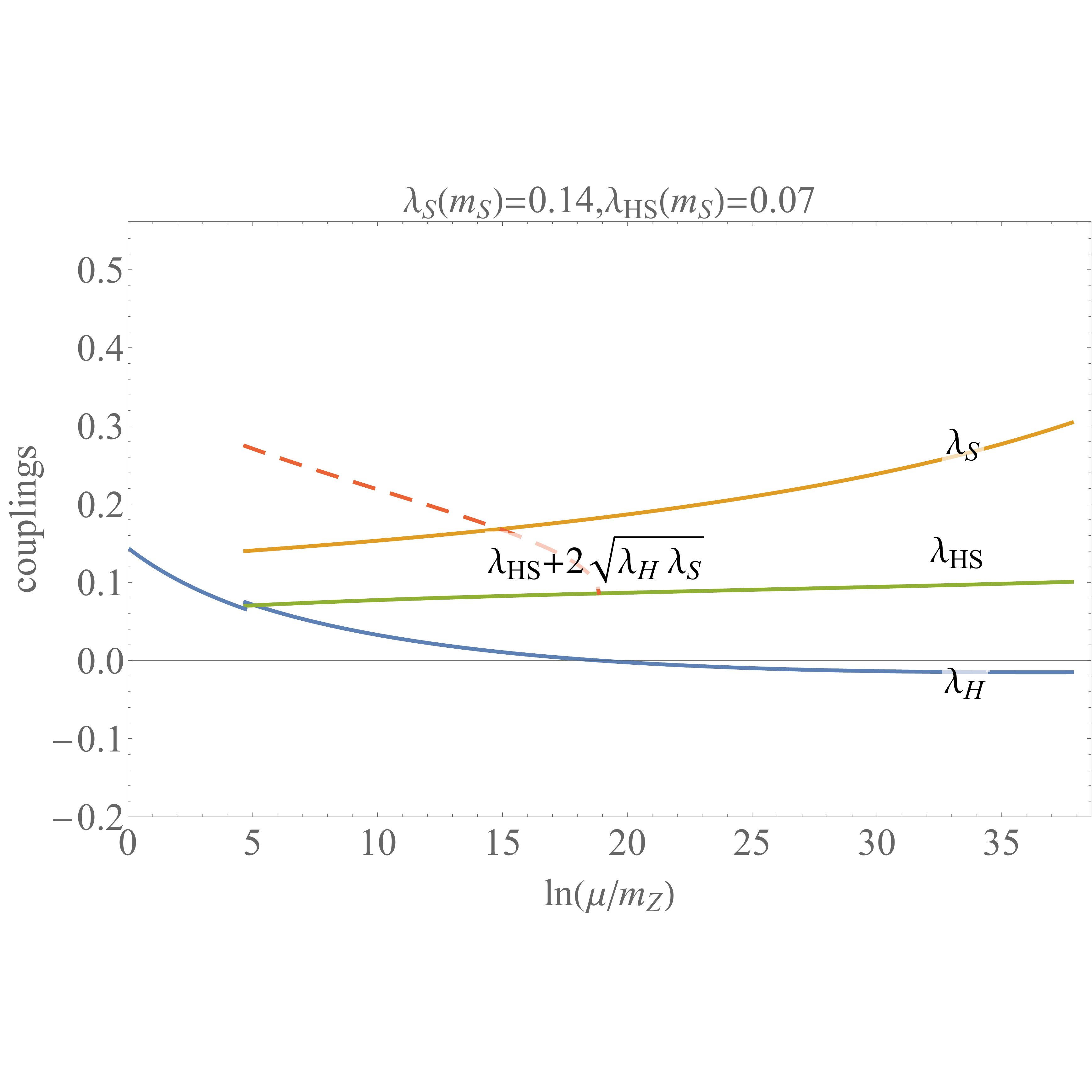}}
	\subfigure[\label{fig5-8}]
	{\includegraphics[width=0.32\textwidth]{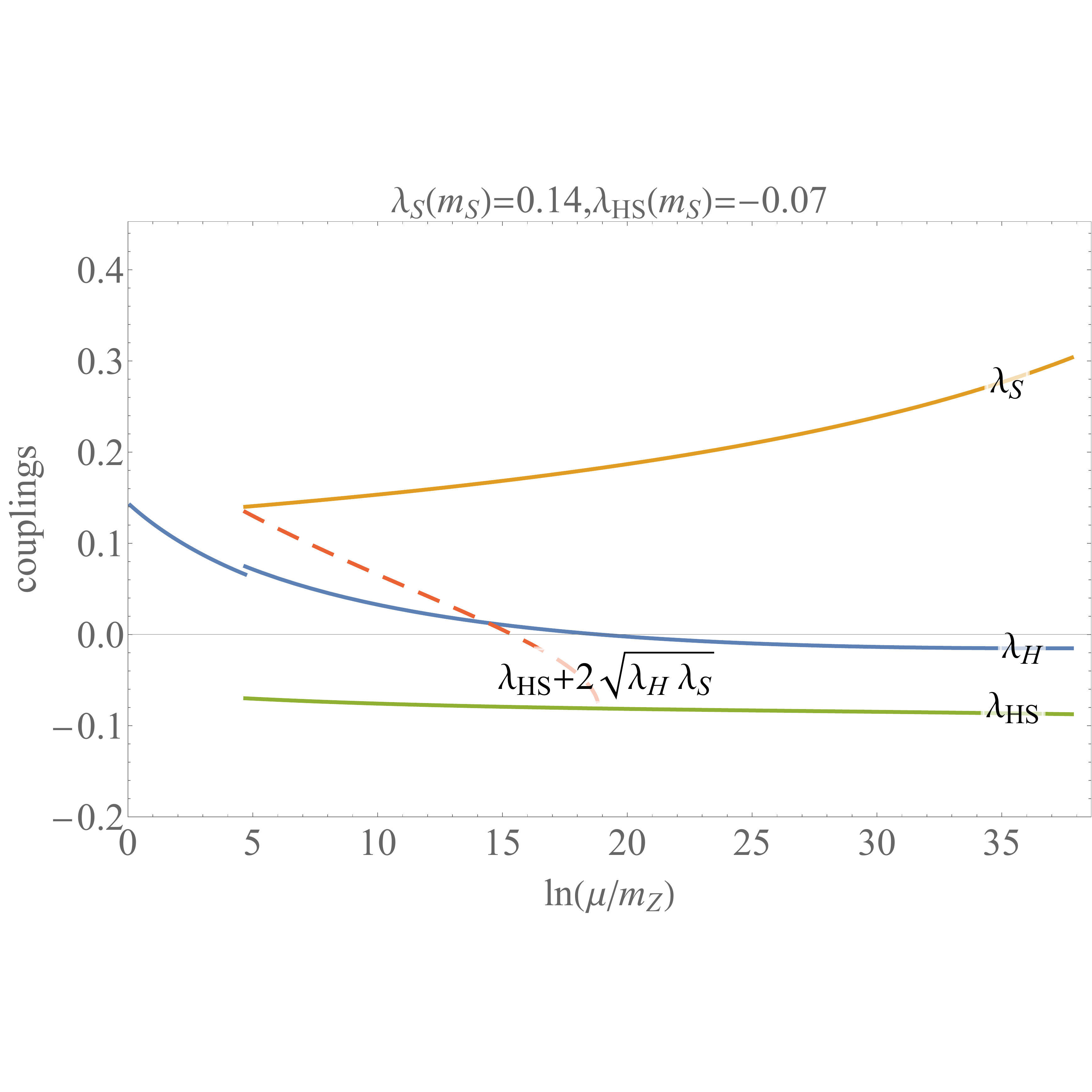}}
	\subfigure[\label{fig5-9}]
	{\includegraphics[width=0.32\textwidth]{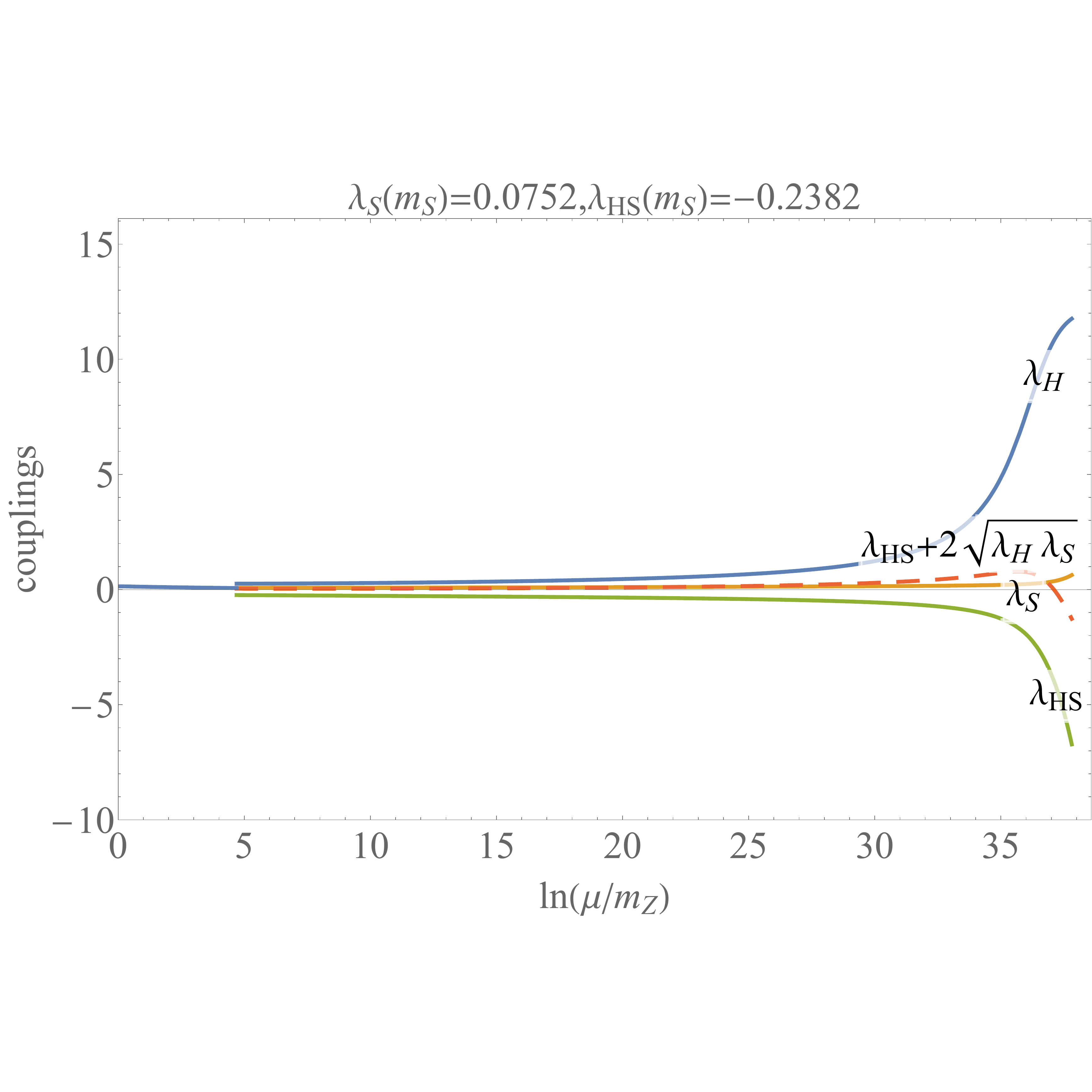}}
	\caption{Evolutions of $\lambda_{H}(\mu)$, $\lambda_{S}(\mu)$, $\lambda_{HS}(\mu)$ with different initial values.}
	\label{running_fig}
\end{figure}

In FIG.\ref{running_fig}, we show the evolutions of quartic couplings in a scale range of $m_Z<\mu<M_P$. The initial values as input for each panel are labeled on the top of each panel. We can see a jump of $\lambda_H$ at $\mu=m_S =10$~TeV is caused by the threshold effect in every plots. The plots shown in the first line corresponds to different choices of parameters with a positive $\lambda_{HS}(m_S)$. From FIGs.\ref{fig5-1} and \ref{fig5-2}, we can see that a threshold correction $\delta\lambda_H\sim\lambda_H$ is helpful for keeping $\lambda_H(\mu)$ (blue lines) far away from the zero during the whole running. FIG.\ref{fig5-3} shows a relatively mild threshold correction and then $\lambda_H(\mu)$ is running towards the zero in high energy scale, but it still remains positive below the Planck scale. As a comparison, FIGs.\ref{fig5-7} and \ref{fig5-8} correspond to the cases that $\lambda_H(\mu)$ becoming negative at some scale lower than $M_P$. The plots shown in the second line corresponds to some chosen parameters with a negative $\lambda_{HS}(m_S)$. In these cases, we should check not only the positivity of $\lambda_H(\mu)$ and $\lambda_S(\mu)$ but also the positivity of the combination $\lambda_{HS}(\mu)+2\sqrt{\lambda_H(\mu)\lambda_S(\mu)}$. For these chosen parameters, this combination keeps positive all the way to the Planck scale, therefore the vacuum stability is ensured. As a comparison,  FIG.\ref{fig5-9}  show an example of $\lambda_{H}(\mu)$ and $\lambda_{S}(\mu)$ keeping positive but $\lambda_{HS}(\mu)+2\sqrt{\lambda_H(\mu)\lambda_S(\mu)}$ becoming negative during the running.

\begin{figure}[!t]
	\centering	
	{\includegraphics[width=0.7\textwidth]{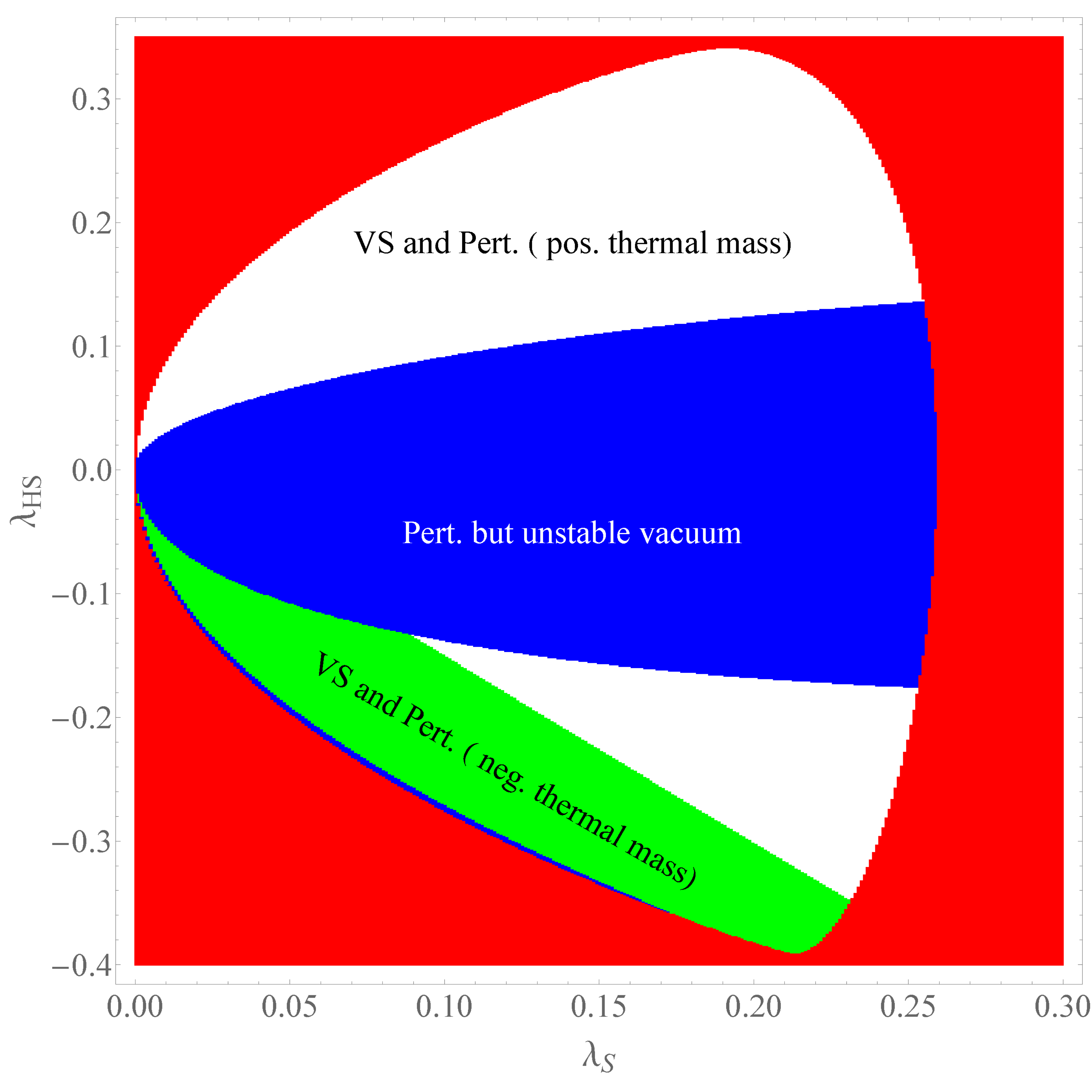}}
	\caption{ The constraints on the $\lambda_S(m_S)$ vs. $\lambda_{HS}(m_S)$ plane from the vacuum stability (VS) and perturbativity conditions up to the Planck scale. The threshold scale is chosen to be $m_S = 10$~TeV. The region shaded in red is excluded by the perturbativity condition, while the region shaded in blue is excluded by the vacuum stability conditions. The green region corresponds to negative thermal corrections to the mass-squared of $H$ and $S$.}
	\label{lamHS-lamS}
\end{figure}

In FIG.\ref{lamHS-lamS}, we show the result of a scan on the $\lambda_S(m_S)$ vs. $\lambda_{HS}(m_S)$ plane, which are input parameters at the threshold scale $m_S=10$~TeV. The values of the SM couplings at the threshold scale can be determined by evolving them from the electroweak scale to $m_S$ by solving the SM RGEs. The region shaded in red corresponds to couplings which become non-perturbative (violating Eq.\eqref{pertcond}) at some scale $\mu_{NP}<M_P$. We can see that it restricts $\lambda_S(m_S)$ and $\lambda_{HS}(m_S)$ in a range $0<\lambda_S(m_S)\lesssim0.25$ and $-0.39\lesssim\lambda_{HS}\lesssim 0.34$. The region shaded in blue is excluded due to violation the vacuum stability (VS) conditions given by Eq.\eqref{VScond}. The region shaded in green corresponds to parameters which lead to negative thermal mass-squared to $H$ or $S$ fields. Negative mass-squared means that the vacuum configuration we have chosen is not stable, so the EW gauge symmetry and the $Z_2$ symmetry are broken rather than restored in the high temperature. For simplicity, we will not operate in this parameter region in this work. We only focus on the blank region in our later discussion on the freeze-in production of DM.

%%%%%%%%%%%%%%

\section{Freeze-in Production of Dark Matter} \label{se.freezein}

\subsection{The Boltzmann Equation}
After the reheating happened, the universe was enclosed by a hot plasma bath of SM and $S$ particles. The dark sector could be produced via freeze-in production due to a feeble Higgs-portal coupling to the SM+$S$ sector. When the reheating temperature of the universe was higher than the mass of the heavy mediator $T_R\gg m_\phi $, the final abundance of DM would be independently of $T_R$. This case is the so-called IR freeze-in production~\cite{Hall10, Abe20}. On the other hand, if $m_\phi \gg T_R $, the relic abundance of DM is determined by the portal coupling $\lambda_{H\Phi}$, $\lambda_{S\Phi}$ and the reheating temperature $T_R$, which is called UV freeze-in~\cite{Hall10,Elahi:2014fsa}.
In this work, we only focus on the IR freeze-in scenario for simplicity and leave the UV freeze-in for our future research.

During the freeze-in production stage, both the dark matter $\chi$ and the CP-even scalar $\phi$ never thermalize. The dominant production processes are similar to the model in Ref.\cite{Abe20}, but there are extra processes such as $S+S\leftrightarrow\phi$, $S+S\leftrightarrow\phi+\phi$, and $S+S\leftrightarrow\chi+\chi$ should be included in our model. To be precise, the Boltzmann equations for $\phi$ and $\chi$ are given by
\begin{eqnarray}
\frac{d n_\phi}{dt} + 3 H n_\phi &=& C_{ \overline{H} H  \leftrightarrow \phi \phi} +C_{\chi \chi \leftrightarrow \phi \phi}+C_{ \overline{H} H  \leftrightarrow \phi } +C_{\chi \chi\leftrightarrow \phi}+C_{S S\leftrightarrow \phi \phi}+C_{S S\leftrightarrow \phi },\nonumber \\
\frac{d n_\chi}{dt} + 3 H n_\chi&=&C_{ \overline{H} H  \leftrightarrow \chi \chi} +C_{\phi \phi \leftrightarrow \chi \chi} +C_{\phi\leftrightarrow \chi \chi }+C_{S S\leftrightarrow \chi \chi},
\end{eqnarray}
where $H = \sqrt{  \pi^2g_\ast/90}T^2/M_P$ is the Hubble parameter during the radiation dominant era, and $M_P=1 / \sqrt{8 \pi G_N}\approx2.4 \times 10^{18}$~GeV is the reduced Planck mass. $C_{A \leftrightarrow B}$ denotes the collision terms for the process $A\leftrightarrow B$ and their definitions can be found in Ref.~\cite{Abe20}, so we just repeat them as follows,
\begin{eqnarray}
C_{i j \ldots \leftrightarrow a b \ldots .}&=& \int \prod_i d \Pi_i f_i \prod_a d \Pi_a\left(1+f_a\right)(2 \pi)^4 \delta^4\left(\sum_i p_i-\sum_a p_a\right) \left|\mathcal{M}_{i j \ldots \rightarrow a b \ldots }\right|^2 \nonumber\\
&&-\int \prod_a d \Pi_a f_a \prod_i d \Pi_i\left(1+f_i\right)(2 \pi)^4 \delta^4\left(\sum_a p_a-\sum_i p_i\right)\left|\mathcal{M}_{a b \ldots \rightarrow i j \ldots}\right|^2,
\end{eqnarray}
where $f_{i,a}$ are distribution function of the initial and final states, and $d\Pi_i=d^3p_i/[(2\pi)^32E_i]$. Since $H$ and $S$ couple to the thermal bath with couplings of $\sim \mathcal{O}(0.01)-\mathcal{O}(0.1)$, they are in thermal equilibrium with the other SM particles during the freeze-in era, and thus their distribution functions are given by the equilibrium ones, $f_H=f_H^{eq}$, $f_S=f_S^{eq}$. On the other hand, due to the feebly coupling, the distribution function of $\phi$ and $\chi$ are assumed to be negligible at order $\mathcal{O}(f_{\phi, \chi}^2)$ in the collision terms.
Therefore, when $m_\phi>2m_H^{(T)},2m_S^{(T)},2m_\chi$, the Boltzmann equations can be written as
\begin{eqnarray}\label{Boltzmann}
\frac{d n_\phi}{dt} + 3 H n_\phi &\approx& C_{ \overline{H} H  \to \phi \phi} +C_{ \overline{H} H  \to \phi }+C_{S S\to \phi \phi}+C_{S S\to \phi }\nonumber\\
&&-(C_{\phi\to \overline{H} H}+C_{\phi\to S S}+C_{\phi\to \chi \chi}),\nonumber\\
\frac{d n_\chi}{dt} + 3 H n_\chi&\approx&C_{ \overline{H} H  \to \chi \chi}^{(subtract)}  +C_{\phi\to \chi \chi }+C_{S S\to \chi \chi}^{(subtract)},
\end{eqnarray}
where the subtracted collision terms $C_{ij\to ab}^{(subtract)}$ in Eq.\eqref{Boltzmann} are defined by subtracting the amplitude squared $|M_{ij\to ab}|^2$ with the one corresponding an on-shell $\phi$ is generated. We will show their precise definition later soon. The decay terms $C_{\phi\to \overline{H} H, SS, \chi\chi}$ can be expressed by
\begin{eqnarray}
C_{\phi\to \overline{H} H, SS, \chi\chi}=\int \frac{d^3 p}{(2\pi)^3} f_\phi \frac{m_\phi}{E}\Gamma_{\phi \to \overline{H} H, SS, \chi\chi },
\end{eqnarray}
and the decay width in c.m. frame are given by
\begin{eqnarray}
\Gamma_{\phi \to \overline{H} H}&=&\frac{\lambda_{H \Phi}^2 m_\phi}{8 \pi \lambda_{\Phi}}\sqrt{1-\frac{4 \left(m_H^{(T)}\right)^2}{m_\phi^2}}~,\nonumber\\
\Gamma_{\phi \to S S}&=&\frac{\lambda_{\Phi S}^2  m_\phi}{32 \pi \lambda_\Phi }\sqrt{1 -\frac{4 \left(m_S^{(T)}\right)^2}{m_\phi^2}}~,\\
\Gamma_{\phi \to \chi\chi}&=&\frac{\lambda_{\Phi} m_\phi}{32 \pi } \sqrt{1-\frac{4 m_\chi^2}{m_\phi^2}}~,\nonumber
\end{eqnarray}
where the masses $\left(m_H^{(T)}\right)^2$ and $\left(m_S^{(T)}\right)^2$ are dominated by the finite temperature corrections: $\left(m_{H,S}^{(T)}\right)^2\approx \delta_Tm_{H,S}^2$ when $T\gg v,w$. Note that $\Gamma_{\phi \to \overline{H}H}$ and $\Gamma_{\phi \to \chi\chi}$ are the same as Eq.(3.7) in Ref.~\cite{Abe20}. For convenience, we can define the branching ratio of the decay process $\phi\to\chi+\chi$ as
\begin{eqnarray}
\mathrm{Br}(\phi\to\chi\chi)\equiv\frac{\Gamma_{\phi\to\chi\chi}}{\Gamma_{\phi\to\chi\chi}+\Gamma_{\phi\to \overline{H} H}+\Gamma_{\phi\to S S}}.
\end{eqnarray}

The collision terms for the inverse decay processes $H^\dag +H\to \phi$ and $S+S \to \phi$ are given by
\begin{eqnarray}
C_{\overline{H} H\to\phi}&=&\sum_{i=1}^2\langle\sigma_{\overline{H_i} H_i \to \phi } \bar v\rangle(n_H^{eq})^2=\frac{m_\phi^2 \Gamma_{\phi \rightarrow \overline{H}H}}{2\pi^2}TK_1(m_\phi/T)~,\\
C_{S S\to\phi}&=&\langle\sigma_{S S \to \phi } \bar v\rangle(n_{s}^{eq})^2=\frac{m_\phi^2 \Gamma_{\phi \to S S}}{2\pi^2}TK_1(m_\phi/T)~,
\end{eqnarray}
while $C_{\chi\chi\to\phi}$ is negligible.
For the collision term of $\overline{H}+H\to\chi+\chi$ processes, we should subtract the contribution from the process that a $\phi$ is generated on-shell. To be precise, we define:
\begin{eqnarray}
C_{\overline{H} H\to\chi\chi}^{(subtract)}&\equiv&C_{\overline{H} H\to\chi\chi}^{(full)}-C_{\overline{H} H\to\chi\chi}^{(RIS)},\\
C_{\overline{H} H\to\chi\chi}^{(full)}&=&2\sum_{i=1}^2\langle\sigma_{\overline{H_i} H_i\to\chi\chi}v\rangle (n_H^{eq})^2\nonumber\\
&=&\frac{T^4\lambda_{H\Phi}^2}{128\pi^5}\int_{2\bar{x}_{\chi,H}}^\infty dz\sqrt{z^2-4x_H^2}\sqrt{z^2-4x_\chi^2}\frac{z^4K_1(z)}{(z^2-x_\phi^2)^2+x_\phi^2\gamma_\phi^2},\\
C_{\overline{H} H\to\chi\chi}^{(RIS)}&\equiv&\lim_{\gamma_\phi\to0}C_{\overline{H} H\to\chi\chi}^{(full)}=\frac{T^4\lambda_{H\Phi}^2}{128\pi^5}\int_{2\bar{x}_{\chi,H}}^\infty dz\sqrt{z^2-4x_H^2}\sqrt{z^2-4x_\chi^2}\frac{\pi z^4K_1(z)}{x_\phi\gamma_\phi}\delta(z^2-x_\phi^2)\nonumber\\
&=&\frac{T^4\lambda_{H\Phi}^2x_\phi^2}{256\pi^4\gamma_\phi}\sqrt{x_\phi^2-4x_H^2}\sqrt{x_\phi^2-4x_\chi^2}K_1(x_\phi)\Theta(x_\phi-2\bar{x}_{\chi,H})\nonumber\\
&=&2C_{\overline{H} H\to\phi}\mathrm{Br}(\phi\to\chi\chi)~,\label{RISHH}
\end{eqnarray}
where $x_{i} \equiv m_{i}/T$~$(i=H,S,\phi,\chi)$, $\gamma_\phi \equiv \Gamma_{\phi}/T$, and $\bar{x}_{\chi,H}=\mathrm{max}(x_\chi,x_H)$. Note that $C_{\overline{H} H\to\chi\chi}^{(RIS)}$ corresponds to the process with on-shell $\phi$, which is called real intermediated state (RIS)~\cite{Kolb:1979qa}. Similarly, we can write down the other subtracted collision terms:
\begin{eqnarray}
C_{S S\to\chi\chi}^{(subtract)}&\equiv&C_{S S\to\chi\chi}^{(full)}-C_{S S\to\chi\chi}^{(RIS)}~,\\
C_{S S\to\chi\chi}^{(full)}&=&2\langle\sigma_{S S\to\chi\chi}v\rangle (n_S^{eq})^2\nonumber\\
&=&\frac{T^4\lambda_{\Phi S}^2}{512\pi^5}\int_{2\bar{x}_{\chi,S}}^\infty dz\sqrt{z^2-4x_S^2}\sqrt{z^2-4x_\chi^2}\frac{z^4K_1(z)}{(z^2-x_\phi^2)^2+x_\phi^2\gamma_\phi^2}~,\\
C_{S S\to\chi\chi}^{(RIS)}&\equiv&\lim_{\gamma_\phi\to0}C_{S S\to\chi\chi}^{(full)}=\frac{T^4\lambda_{\Phi S}^2}{512\pi^5}\int_{2\bar{x}_{\chi,S}}^\infty dz\sqrt{z^2-4x_S^2}\sqrt{z^2-4x_\chi^2}\frac{\pi z^4K_1(z)}{x_\phi\gamma_\phi}\delta(z^2-x_\phi^2)\nonumber\\
&=&\frac{T^4\lambda_{\Phi S}^2x_\phi^2}{1024\pi^4\gamma_\phi}\sqrt{x_\phi^2-4x_S^2}\sqrt{x_\phi^2-4x_\chi^2}K_1(x_\phi)\Theta(x_\phi-2\bar{x}_{\chi,S})\nonumber\\
&=&2C_{S S\to\phi}\mathrm{Br}(\phi\to\chi\chi)~,\label{RISSS}
\end{eqnarray}
where $\bar{x}_{\chi,S}=\mathrm{max}(x_\chi,x_S)$.

Since $\phi$ will completely decay into $\chi$ and other particles in the end, we can define the final dark matter number density as~\cite{Abe20}
%%%%%%%%
\begin{eqnarray}
n_D = n_\chi + 2\mathrm{Br}(\phi\to\chi\chi)\times n_\phi~.
\end{eqnarray}
Now we can combine Eqs.\eqref{Boltzmann} and find the equation for $n_D$ as follows
\begin{eqnarray}
\frac{d n_D}{dt} + 3 H n_D&\approx&(C_{\overline{H} H\to\phi\phi}+C_{SS\to\phi\phi})\cdot 2\mathrm{Br}(\phi\to\chi\chi)\nonumber\\
&&+C_{\overline{H} H\to\chi\chi}^{(subtract)}+C_{SS\to\chi\chi}^{(subtract)}+(C_{H^\dag H\to\phi}+C_{SS\to\phi})\cdot 2\mathrm{Br}(\phi\to\chi\chi)\nonumber\\
&&-(C_{\phi\to \overline{H} H}+C_{\phi\to SS}+C_{\phi\to \chi\chi})\cdot 2\mathrm{Br}(\phi\to\chi\chi)+C_{\phi\to\chi\chi}\nonumber\\
&=&C_{\overline{H} H\to\chi\chi}^{(full)}+C_{SS\to\chi\chi}^{(full)}+(C_{\overline{H} H\to\phi\phi}+C_{SS\to\phi\phi})\cdot 2\mathrm{Br}(\phi\to\chi\chi)~.\label{BE}
\end{eqnarray}

\begin{figure}[!t]
	\centering
	\subfigure[\label{figHHphiphiX}]
	{\includegraphics[width=0.23\textwidth]{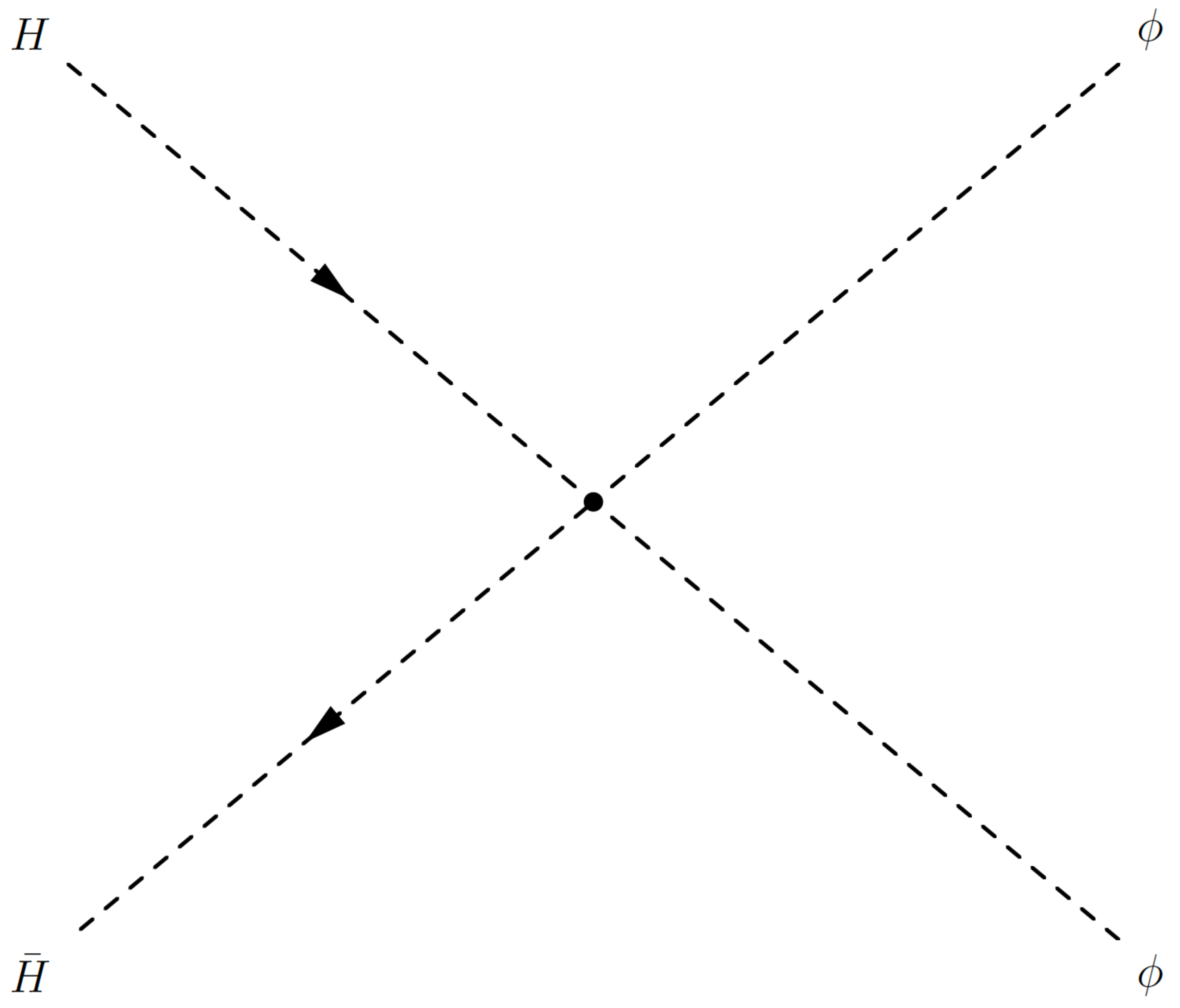}}~
	\subfigure[\label{figHHphiphis}]
	{\includegraphics[width=0.23\textwidth]{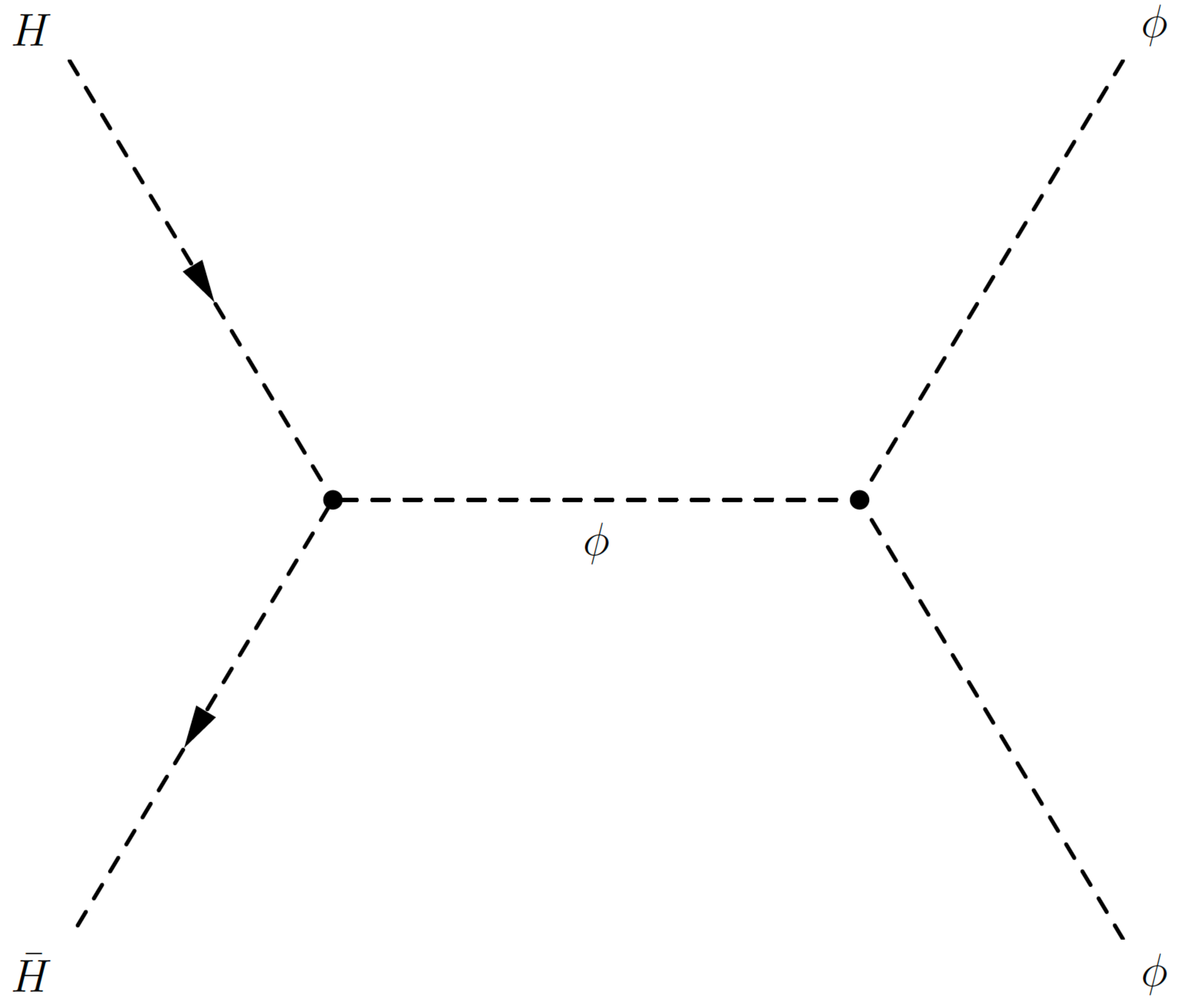}}~
	\subfigure[\label{figHHphiphit}]
	{\includegraphics[width=0.23\textwidth]{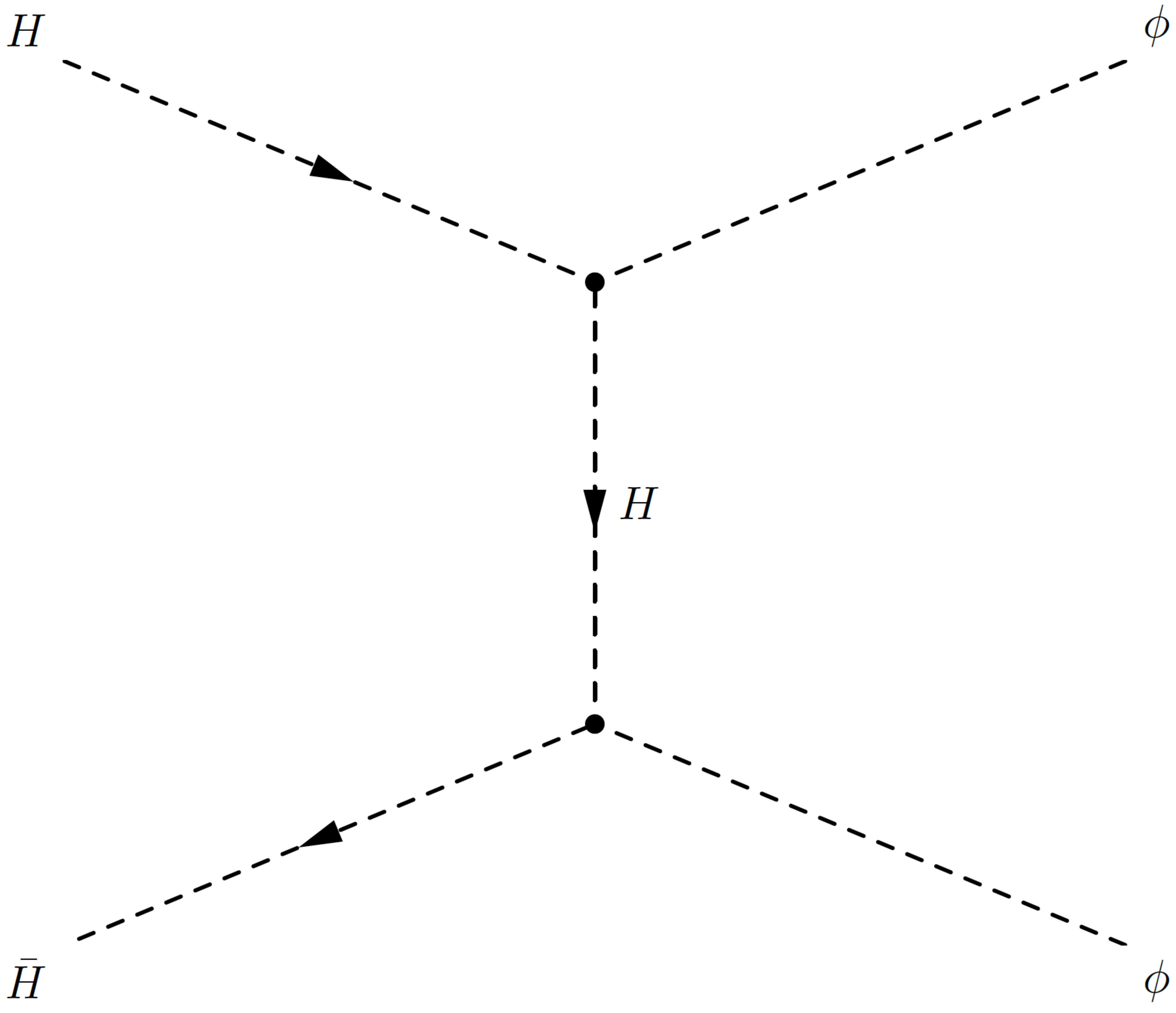}}~
	\subfigure[\label{figHHphiphiu}]
	{\includegraphics[width=0.23\textwidth]{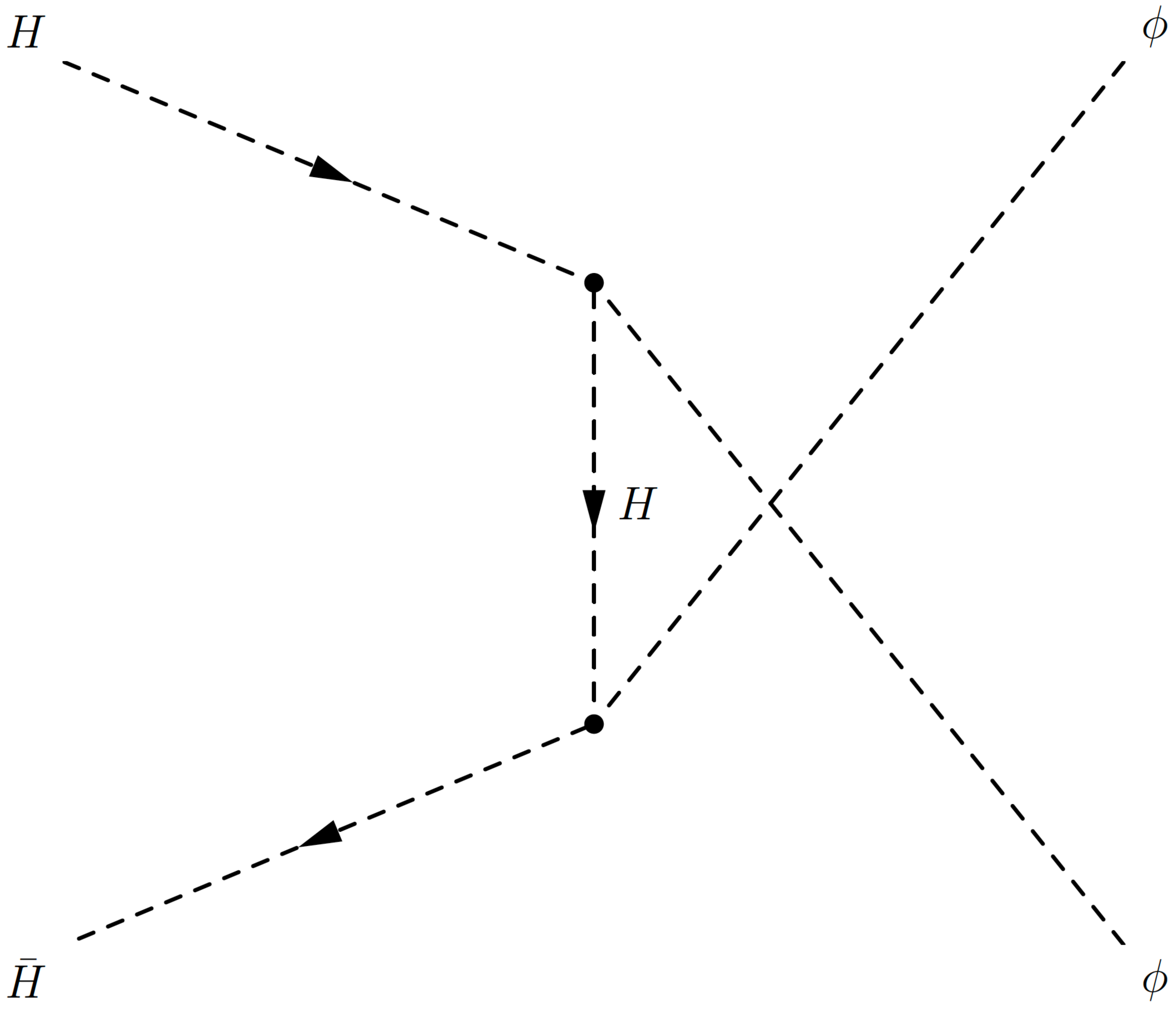}}
	\subfigure[\label{figSSphiphiX}]
	{\includegraphics[width=0.23\textwidth]{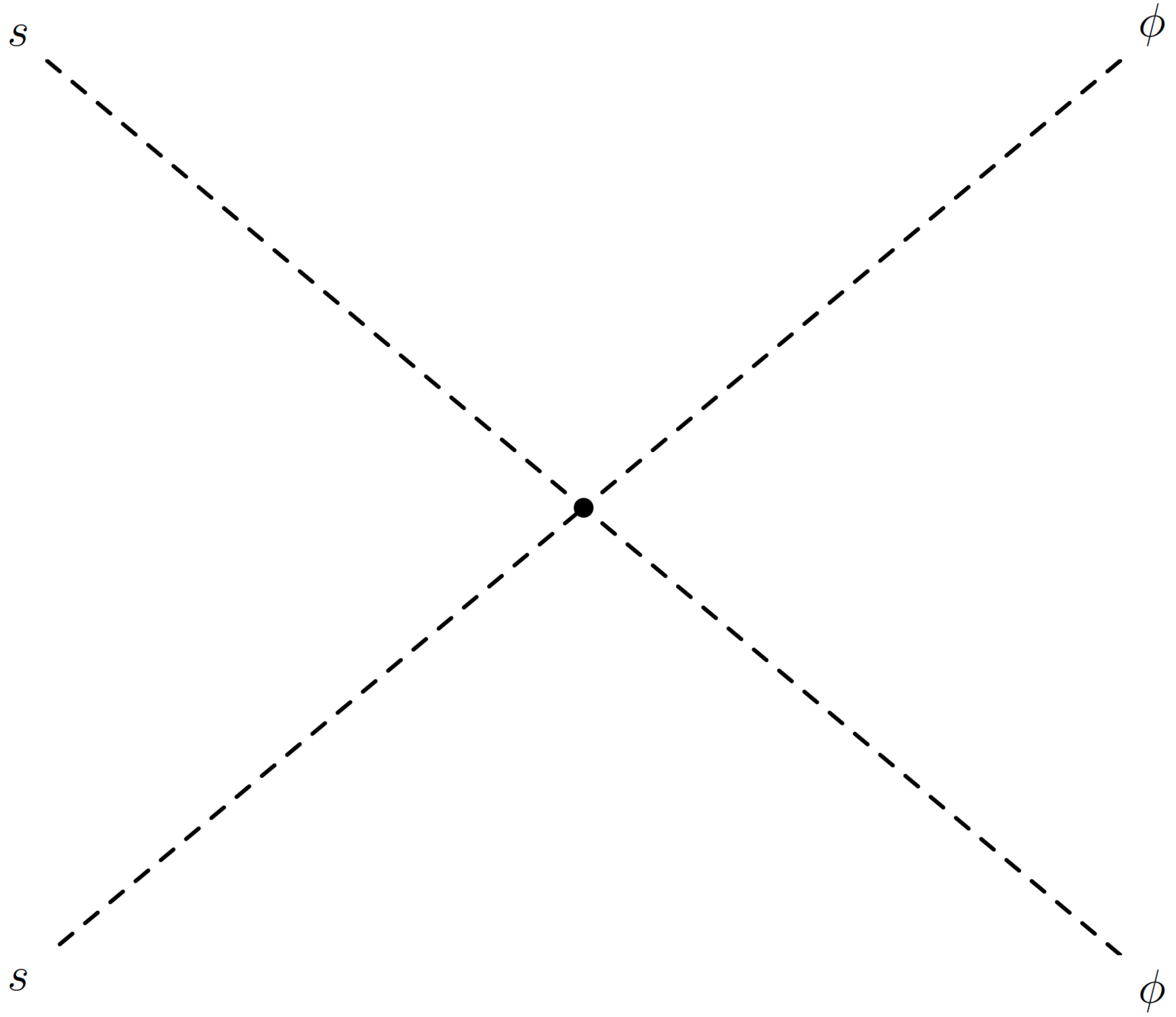}}~
	\subfigure[\label{figSSphiphis}]
	{\includegraphics[width=0.23\textwidth]{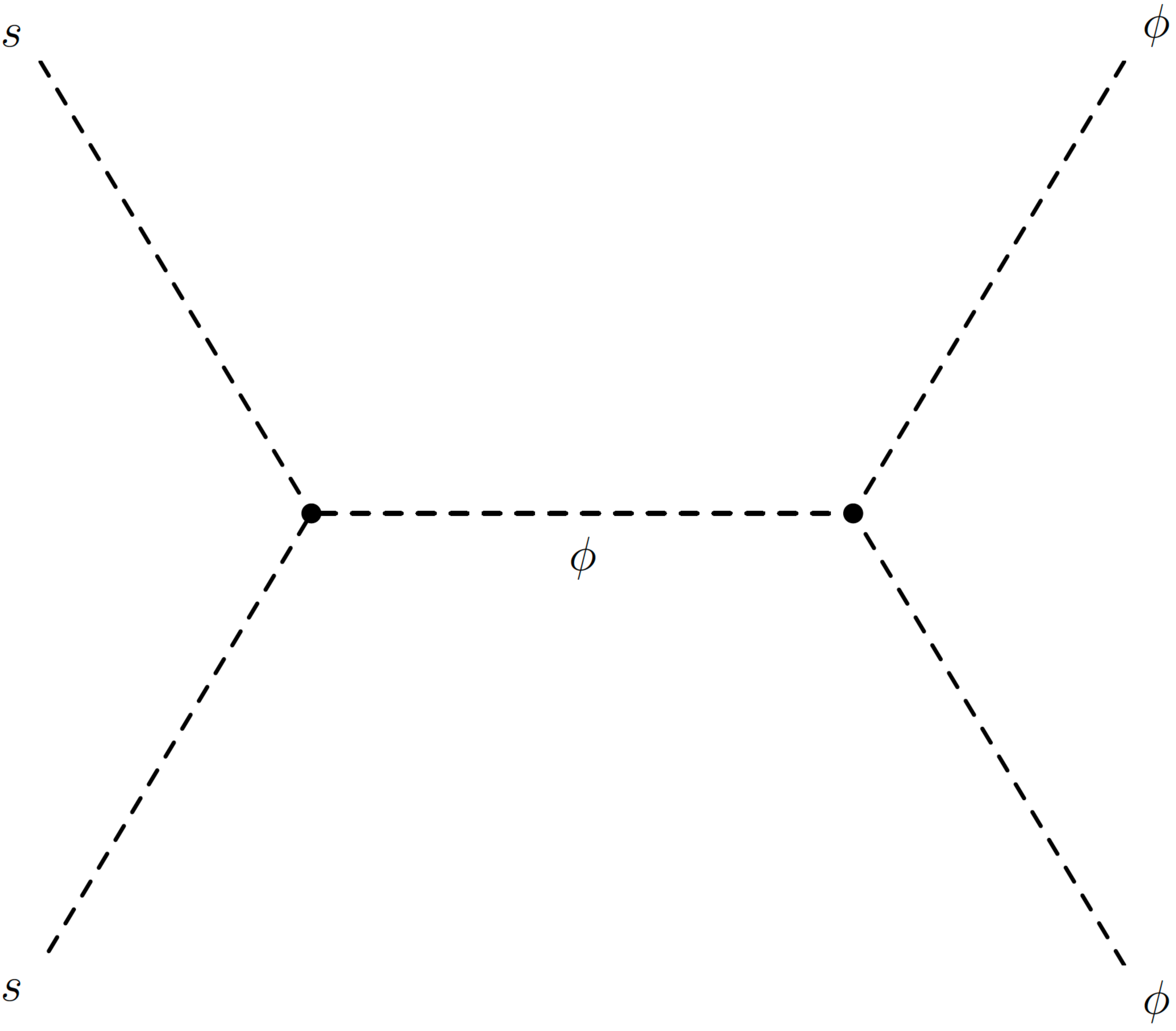}}~
	\subfigure[\label{figSSphiphit}]
	{\includegraphics[width=0.23\textwidth]{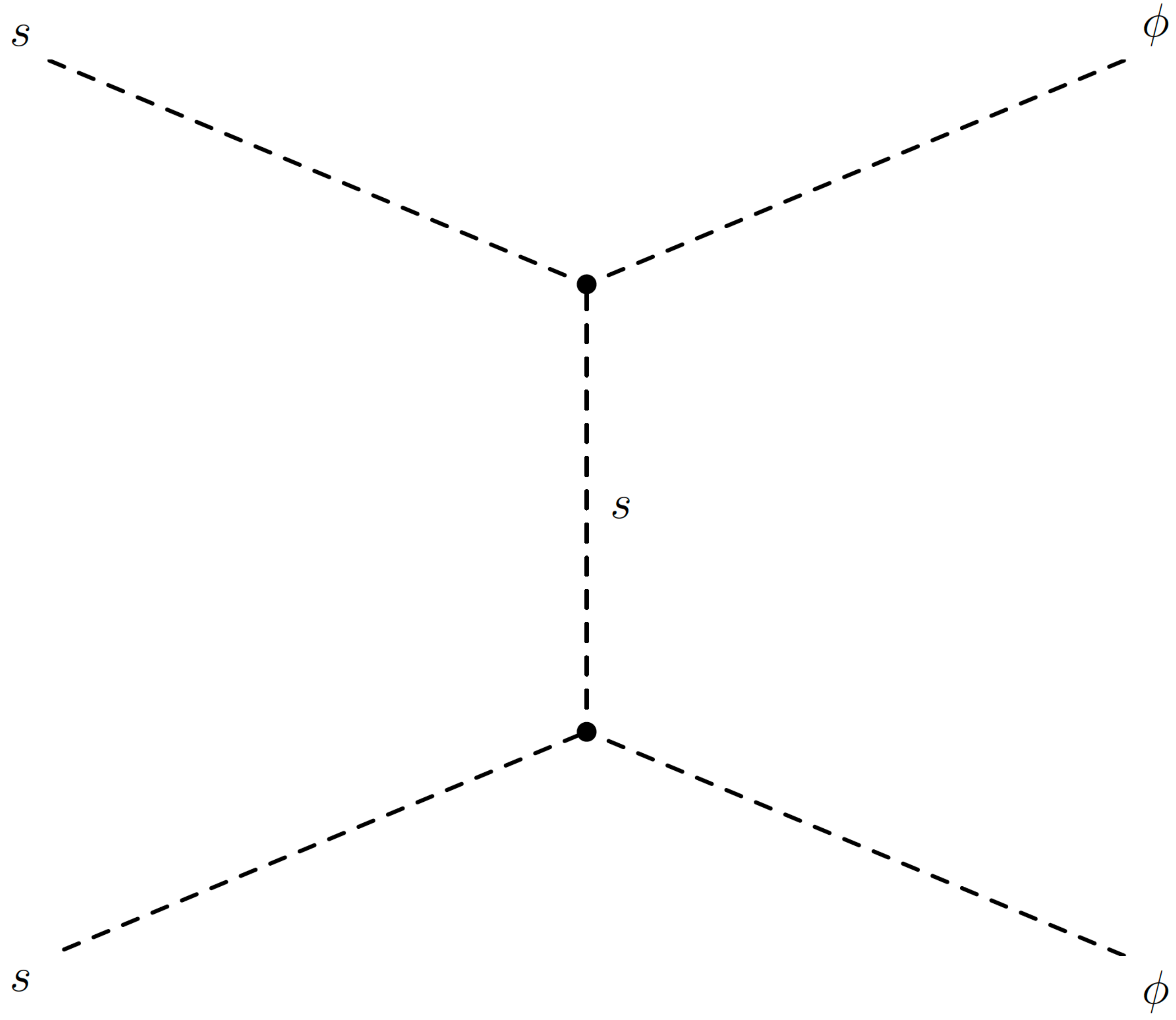}}~
	\subfigure[\label{figSSphiphiu}]
	{\includegraphics[width=0.23\textwidth]{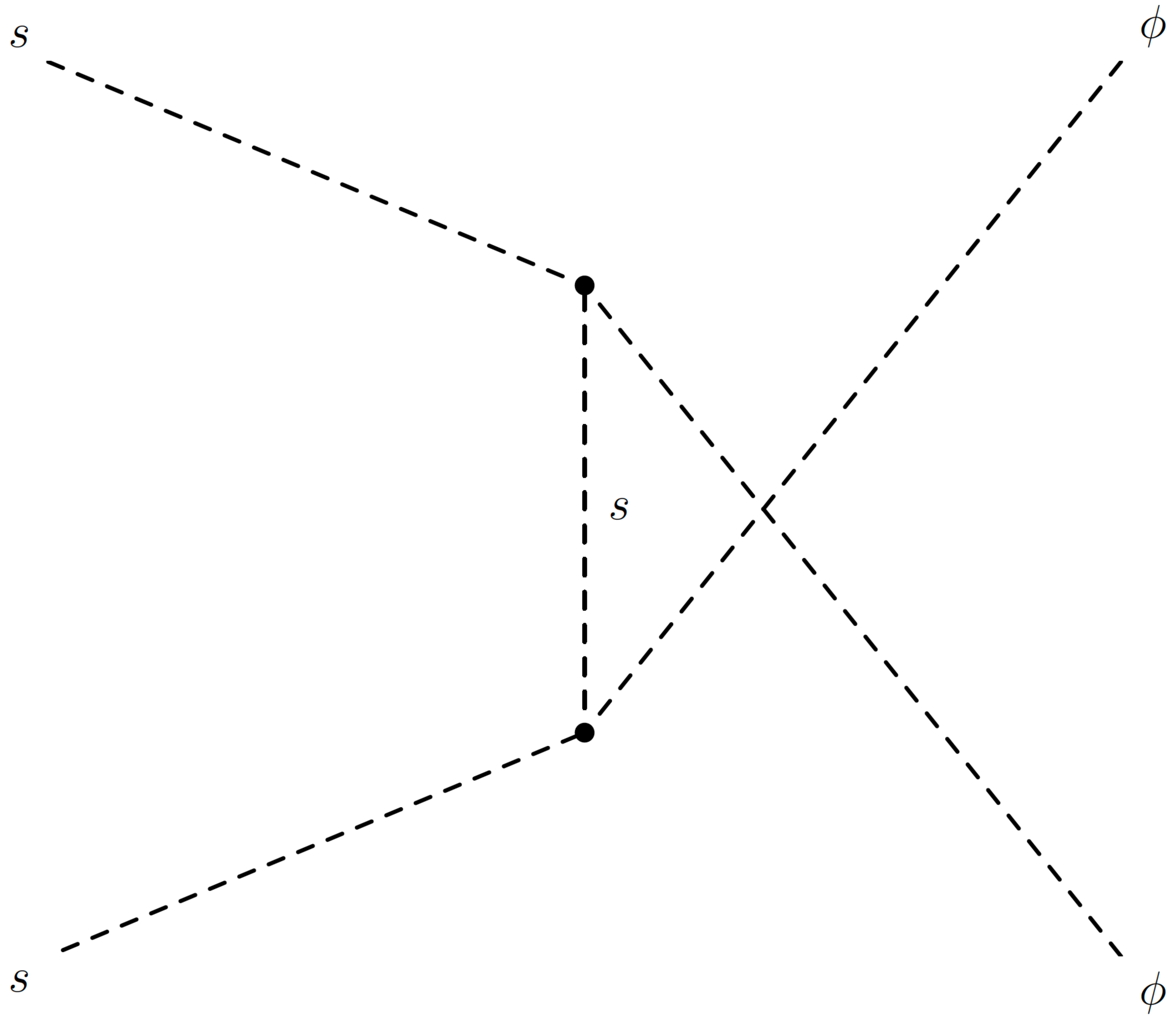}}
	\caption{Feynman diagrams for the $\overline{H}+H\to \phi+\phi$ and $S+S\to \phi+\phi$.}
	\label{Feyndiag}
\end{figure}

The collision term for $\overline{H}+H\to \phi+\phi$ is given by,
\begin{eqnarray}\label{colHHphiphi}
C_{\overline{H} H\to\phi\phi}&=&2\sum_{i=1}^2\langle\sigma_{\overline{H_i} H_i  \rightarrow \phi\phi} \bar v\rangle(n_H^{eq} )^2\nonumber\\
&=&2\times\frac{T^4}{512\pi^5}\int_{2\overline{x}_{\phi,H}}^{\infty}dz ( z^2 -4x_H^2)^{1/2}( z^2- 4x_\phi^2)^{1/2}{K_1(z)}\nonumber \\
&&\times \frac{1}{2}\int_{-1}^{1} d\cos\theta\sum_{i=1}^2|\mathcal{M}_{\overline{H_i} H_i\to \phi\phi}|^2~,
\end{eqnarray}
where $\overline{x}_{\phi}=\mathrm{max}(x_\phi,x_H)$, and the total amplitude squared is given by
\begin{eqnarray}\label{ampsHHphiphi}
\sum_{i=1}^2|{\cal M}_{\overline{H_i} H_i\to \phi\phi }|^2=\sum_{i=1}^2|{\cal M}_{\overline{H_i} H_i\to \phi\phi,4}+{\cal M}_{\overline{H_i} H_i \to \phi\phi,s}+{\cal M}_{\overline{H_i} H_i \to \phi\phi,t}+{\cal M}_{\overline{H_i} H_i\to \phi\phi,u}|^2~.
\end{eqnarray}
The partial amplitudes for different channels can be derived as
\begin{eqnarray}  \label{ampHHphiphi}
&&i{\cal M}_{\overline{H_i} H_i\to \phi \phi,4}= -i\lambda_{H \phi},\quad i{\cal M}_{\overline{H_i} H_i\to \phi \phi,s} =-i\frac{3 \lambda_{H\phi}x_\phi^2}{(z^2-x^2_{\phi} + i x_\phi \gamma_\phi)}~,\nonumber \\
&&i{\cal M}_{\overline{H_i} H_i\to \phi \phi,t}=-i\frac{(\lambda_{H\Phi}^2/\lambda_\Phi) x_\phi^2}{x_\phi^2 -\frac{1}{2}z^2(1-\beta_\phi \beta_H \cos\theta)}~,\nonumber\\
&&i{\cal M}_{\overline{H_i} H_i\to \phi \phi,u}=-i\frac{(\lambda_{H\phi}^2/\lambda_\Phi) x_\phi^2}{x_\phi^2 -\frac{1}{2}z^2(1+\beta_\phi \beta_H \cos\theta)}~,
\end{eqnarray}
where $\beta_{\phi,H}=\sqrt{1 -4x_{\phi,H}^2/z^2} $. In the first line of FIG.\ref{Feyndiag}, we show the Feynman diagrams for these channels.

%%%%%%%%%%%%%%%%
Similarly, we can find the collision terms for $S+S\to \phi+\phi$:
\begin{eqnarray}\label{colSSphiphi}
C_{S S\to\phi\phi}&=&2\langle\sigma_{S S  \rightarrow \phi} \bar v\rangle(n_{S}^{eq})^2\nonumber\\ &=&2\times\frac{1}{2}\times\frac{T^4}{512\pi^5}\int_{2\overline{x}_{\phi,S}}^{\infty}dz [ z^2 -4x_S^2]^{1/2}[ z^2- 4x_\phi^2]^{1/2}{K_1(z)}\nonumber \\
&&\qquad\times\frac{1}{2}\int d\cos \theta|\mathcal{M}|^2_{SS\to \phi\phi}~,
\end{eqnarray}
where the total amplitude-squared is given by
\begin{eqnarray}\label{ampsSSphiphi}
|{\cal M}_{S S\to \phi\phi }|^2=|{\cal M}_{S S\to \phi\phi,4}+{\cal M}_{S S\to \phi\phi,s}+{\cal M}_{S S \to \phi\phi,t}+{\cal M}_{S S\to \phi\phi,u}|^2~,
\end{eqnarray}
with the partial amplitudes:
\begin{eqnarray}  \label{ampSSphiphi}
&&i{\cal M}_{S S \to \phi \phi,4}= -i\lambda_{\Phi S}~,\quad i{\cal M}_{S S \to \phi \phi,s} =-i\frac{3\lambda_{\Phi S}x_\phi^2}{(z^2-x^2_{\phi} + i x_\phi \gamma_\phi)}~,\nonumber \\
&&i{\cal M}_{S S \to \phi \phi,t} =-i\frac{(\lambda_{\Phi S}^2/\lambda_\Phi) x_\phi^2}{x_\phi^2 -\frac{1}{2}z^2(1-\beta_\phi \beta_S \cos\theta)}~,\nonumber \\
&&i{\cal M}_{S S \to \phi \phi,u}=-i\frac{(\lambda_{\Phi S}^2/\lambda_\Phi) x_\phi^2}{x_\phi^2 -\frac{1}{2}z^2(1+\beta_\phi \beta_S \cos\theta)}~,
\end{eqnarray}
where $\beta_{S}=\sqrt{1 -4x_{S}^2/z^2} $. The corresponding Feynman diagrams are shown in the second line of FIG.\ref{Feyndiag}.

Note that if we drop all the terms involving $S$ field, and ignore the amplitudes corresponding to t- and u-channels of the process $\overline{H}+H \to \phi+\phi$, we reach the same results shown in Eq.(3.10)-(3.12) of Ref.\cite{Abe20}. This is a proper approximation when $\lambda_{H\Phi}^2\lesssim \lambda_\Phi^2$. However, in a situation that $\lambda_{H\Phi}^2\gg \lambda_\Phi^2$, these two channels might not be negligible. We compare the results with and without t- and u-channels in appendix~\ref{app1}.

For solving the Boltzmann equation Eq.\eqref{BE}, we define the yield of the dark matter, $Y_D$, as follows,
\begin{eqnarray}
Y_D=\frac{n_D}{s},
\end{eqnarray}
where $s=2\pi^2g_{\ast s}T^3/45$ is the entropy density, and $g_{\ast s}\approx g_\ast$ is the effective relativistic degree of freedom of particles in the thermal bath. The left hand side of Eq.\eqref{BE} can now be rewritten in terms of $Y_D$ as
\begin{eqnarray}
\frac{dn_D}{dt}+3 H n_D=-H T s\frac{dY_D}{dT}~,
\end{eqnarray}
and then Eq.\eqref{BE} can be solved by integration
\begin{eqnarray}\label{YD}
Y_D&\approx&-\int_{T_R}^{0} \frac{dT}{sH T }\left[C_{\overline{H} H\to\chi\chi}^{(full)}+C_{SS\to\chi\chi}^{(full)}+(C_{\overline{H} H\to\phi\phi}+C_{SS\to\phi\phi})\cdot 2\mathrm{Br}(\phi\to\chi\chi)\right]~.
\end{eqnarray}

Note that if $\lambda_{H\Phi}^2,~\lambda_{\Phi S}^2\lesssim\lambda_\Phi^2$, the dominant contributions in the integrand of Eq.\eqref{YD} is the RIS parts, Eq.\eqref{RISHH} and \eqref{RISSS}. In this situation, we can approximate $Y_D$ by
\begin{eqnarray}\label{approxYD}
Y_D&\approx&-\int_{T_R}^{0} \frac{dT}{sH T }\left[C_{\overline{H} H\to\chi\chi}^{(RIS)}+C_{SS\to\chi\chi}^{(RIS)}\right]\nonumber\\
&=&-\int_{T_R}^{0} \frac{dT}{sH T }\left(C_{\overline{H} H\to\phi}+C_{S S\to\phi}\right)\cdot 2\mathrm{Br}(\phi\to\chi\chi)\nonumber\\
&=&-\int_{T_R}^{0} \frac{dT}{sH T }\frac{m_\phi^2}{\pi^2}TK_1\left(\frac{m_\phi}{T}\right)\mathrm{Br}(\phi\to\chi\chi)(\Gamma_{\phi \rightarrow \overline{H}H}+\Gamma_{\phi \rightarrow SS})\nonumber\\
&=&\frac{45M_P}{16\pi^6\lambda_\Phi m_\phi}\sqrt{\frac{g_\ast}{90}}\int_{m_\phi/T_R}^{\infty}dx_\phi x_\phi^3K_1\left(x_\phi\right)\nonumber\\
&&\times\frac{\lambda_\Phi^2\sqrt{1-\left(\frac{2x_\chi}{x_\phi}\right)^2}}{\lambda_\Phi^2\sqrt{1-\left(\frac{2x_\chi}{x_\phi}\right)^2}+4\lambda_{H \Phi}^2\sqrt{1-\left(\frac{ 2x_H}{x_\phi}\right)^2}\Theta\left(x_\phi-2x_{H}\right)+\lambda_{\Phi S}^2\sqrt{1-\left(\frac{ 2x_S}{x_\phi}\right)^2}\Theta\left(x_\phi-2x_{S}\right)}\nonumber\\
&&\times\left[\lambda_{H \Phi}^2\sqrt{1-\left(\frac{ 2x_H}{x_\phi}\right)^2}\Theta\left(x_\phi-2x_{H}\right)+\frac{\lambda_{\Phi S}^2}{4}\sqrt{1-\left(\frac{ 2x_S}{x_\phi}\right)^2}\Theta\left(x_\phi-2x_{S}\right)\right]~.\nonumber\\
\end{eqnarray}
The wave of massive production of DM happens at the time that a real $\phi$ particle can be produced by pair annihilation of $H$ or $S$. In a situation that $\lambda_{H\Phi}^2,~\lambda_{\Phi S}^2\gg\lambda_\Phi^2$, the contributions from $C_{\overline{H} H\to\phi\phi}$ and $C_{SS\to\phi\phi}$ should be included, but the $C_{\overline{H} H\to\phi\phi}^{(full)}$ and  $C_{S S\to\phi\phi}^{(full)}$ terms can still be approximated by their RIS parts. Note that in principle we should treat all the running couplings as functions of temperature by taking $\mu\sim T$, but we find that the collision terms have no significant different if we use the values of couplings at $\mu\sim m_\phi$. It is because the production reaction is only significant at a temperature around $T\approx2m_H^{(T)}\sim m_\phi$.

\begin{figure}[!t]
	\centering
	{\includegraphics[width=0.7\textwidth]{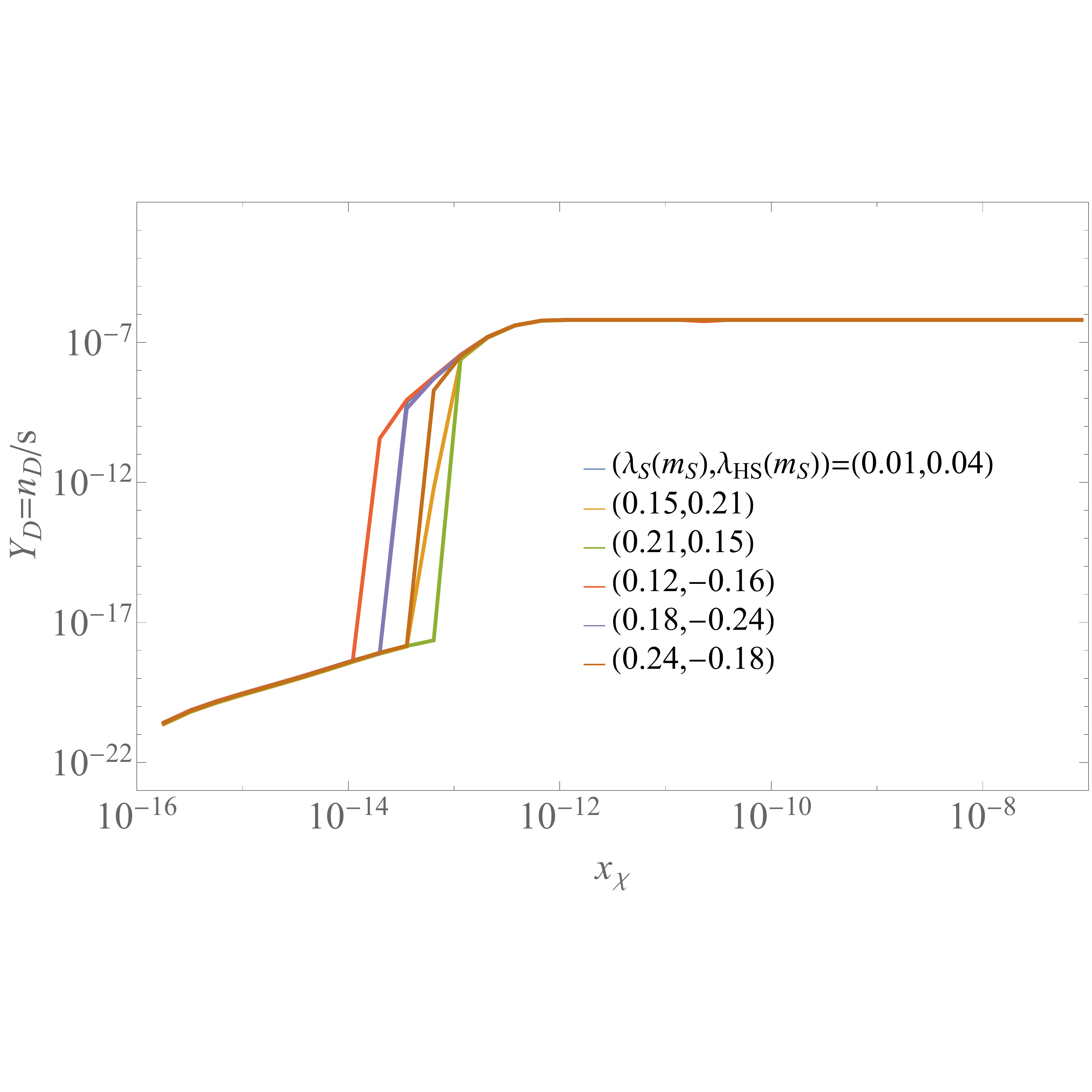}}
	\caption{Yields of DM evolve with $x_\chi\equiv m_\chi/T$ for different $\lambda_S(m_S), \lambda_{ HS}(m_S)$. }
	\label{Yield}
\end{figure}

In FIG.\ref{Yield}, we show the numerical results of evolving $Y_D=n_D/s$ for several different benchmark points of $(\lambda_S(m_S),\lambda_{HS}(m_S))$ at $m_S=10^4$~GeV.
The other parameters in the two panels are chosen in common as follows,
\begin{eqnarray}
\lambda_{H\Phi}=\lambda_{\Phi S}=\lambda_\Phi = 10^{-10},\quad m_\phi = 10^{10}~\text{GeV},\quad m_\chi = 10^{-3}~\text{GeV}.
\end{eqnarray}
We can see that different values of $\lambda_{S}(m_S)$ and $\lambda_{HS}(m_S)$ only shift the temperature of explosive production, and the final yield of DM, $Y_D(\infty)$, converges to the same value. This can be understood by the approximated $Y_D$ given by Eq.\eqref{approxYD}. In the integrand, there are $(2x_{H,S})^2$ given by
\begin{eqnarray}
(2x_H)^2&\approx& \frac{1}{4}g'^2(m_\phi)+\frac{3}{4}g^2(m_\phi)+y_t^2(m_\phi)+\frac{1}{6}\lambda_{HS}(m_\phi)+2\lambda_{H}(m_\phi)\nonumber\\
&\approx& 0.58+\frac{1}{6}\lambda_{HS}(m_\phi)+2\lambda_{H}(m_\phi)~,\\
(2x_S)^2&\approx&\frac{2}{3}\lambda_{HS}(m_\phi)+\lambda_S(m_\phi)~,
\end{eqnarray}
which can not exceed $1$ for $m_\phi\lesssim10^{10}$~GeV, otherwise $\lambda_{S}(\mu)$, $\lambda_{H}(\mu)$ and $\lambda_{HS}(\mu)$ will become nonperturbative below the Planck scale. On the other hand, the maximal value of $x_\phi^3 K_(x_\phi)$ is reached around $x_\phi\approx2.39$, then $(2x_{H,S})/x_\phi^2$ is negligible comparing to $1$, and thus the integration is not sensitive to $\lambda_{S}(m_S)$ and $\lambda_{HS}(m_S)$ in the viable region. They are only sensitive to the portal couplings, $\lambda_{H\Phi},\lambda_{S\Phi}$, which are set to a common value for every cases in FIG.\ref{Yield}, so the final results of $Y_D$ for different benchmark points converge to a same value.

\begin{figure}[!t]
	\centering
	\subfigure[\label{fig7-1}]
	{\includegraphics[width=0.49\textwidth]{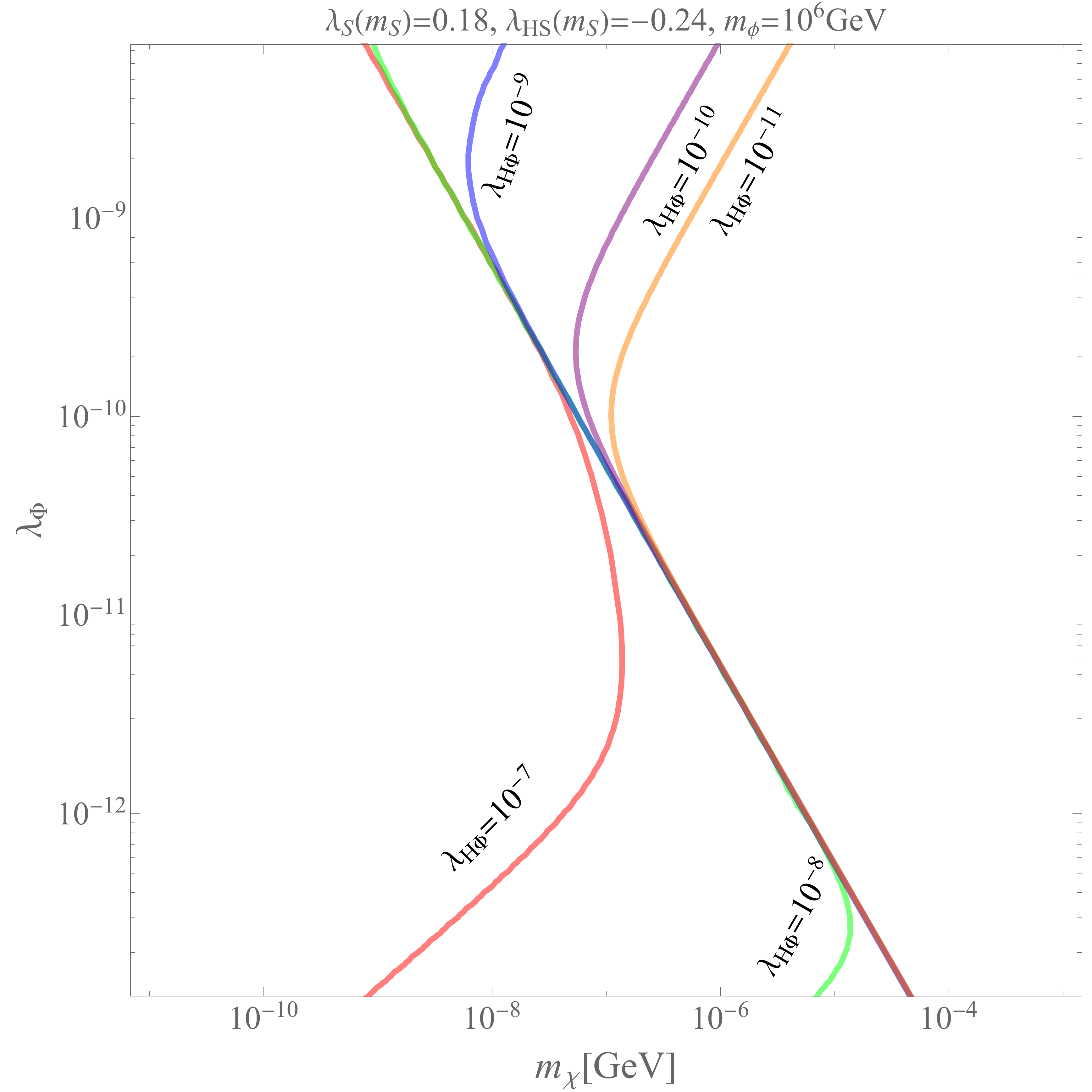}}
	\subfigure[\label{fig7-2}]
	{\includegraphics[width=0.49\textwidth]{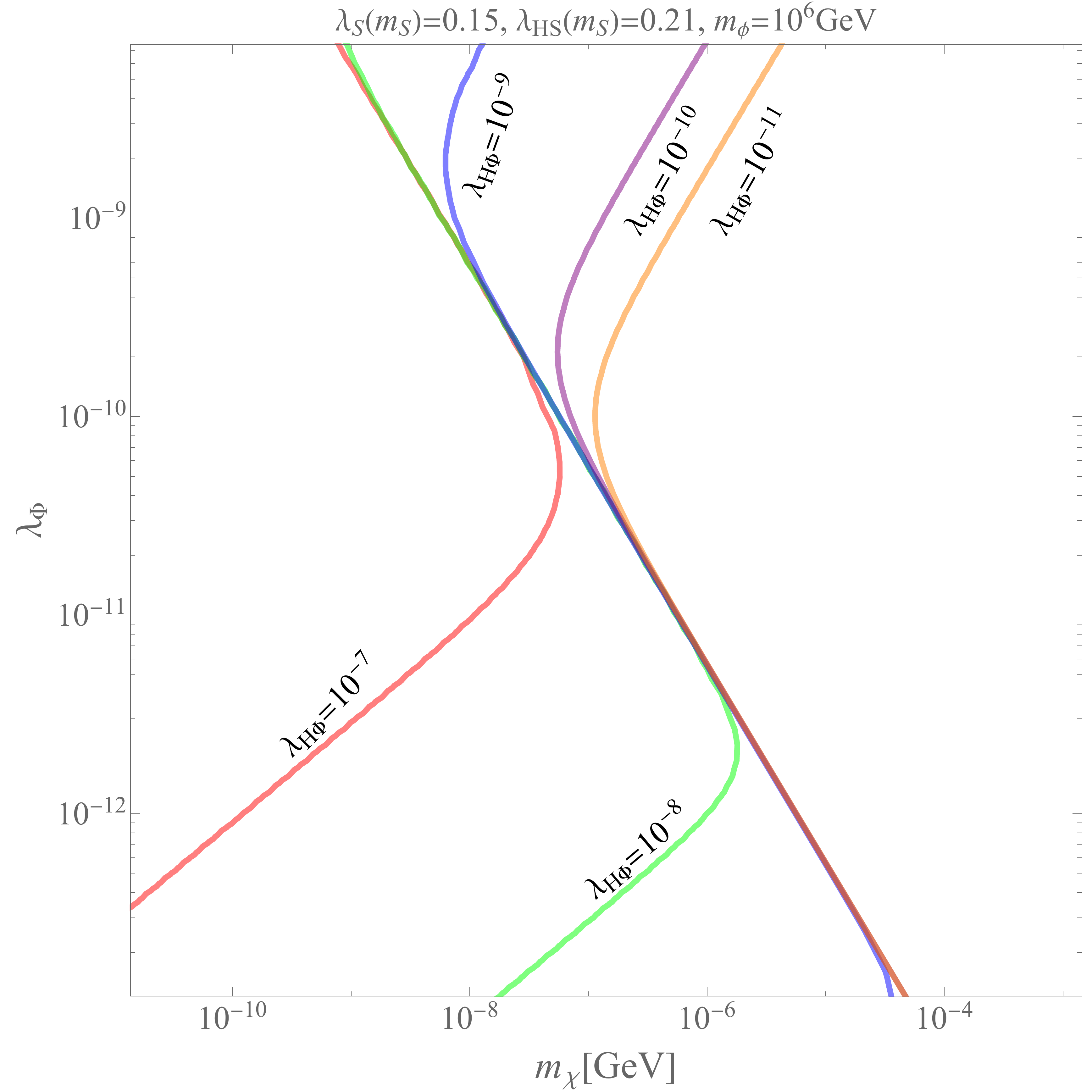}}
	\caption{ Contours that reproduce the observed relic abundance in the $m_\chi$ vs. $\lambda_\Phi$
		plane with fixing $m_\phi = 10^6$~GeV, and $\lambda_{ \Phi S} = 10^{-10}$.
		The red, green, blue, purple, and orange lines correspond to $\lambda_{H\Phi} = 10^{-7},~10^{-8},~10^{ -9},~10^{ -10}$, and $10^{ -11}$ respectively. (Left): $\lambda_S(m_S) = 0.18,~\lambda_{HS}(m_S) = -0.24$. (Right): $\lambda_S(m_S) = 0.15,~\lambda_{HS}(m_S) = 0.21$.
	}
	\label{fig7}
\end{figure}

In FIG.\ref{fig7}, we show some contours in the $m_\chi$ vs. $\lambda_\Phi$ plane, which can reproduce the dark matter relic abundance implied by the observation in the PLANCK experiment~\cite{Planck:2018vyg}.
The parameters are chosen as follows
\begin{eqnarray}
&&\lambda_{H\Phi} = 10^{-11} \sim 10^{ -7}, \quad \lambda_{\Phi S}=\lambda_\Phi = 10^{-10},\quad \lambda_{HS} = -0.24(\textrm{Left}),~0.21(\textrm{Right})~,\nonumber \\
&&\lambda_S = 0.18(\textrm{Left}),~0.15(\textrm{Right}),\quad m_\phi = 10^{6}~\text{GeV},\quad m_\chi = 10^{-3} ~\text{GeV}.
\end{eqnarray}
In both panels, we can see that all the contours have a constant value of $\lambda_\Phi/m_\chi$ in the range corresponding to $\lambda_{H\Phi},\lambda_{\Phi S}\lesssim\lambda_\Phi$. It can be understood according to Eq.\eqref{approxYD}, that in the limit of $\lambda_{H\Phi},\lambda_{\Phi S}\ll\lambda_\Phi$, $\textrm{Br}(\phi\to\chi\chi)\to1$, so the relic abundance has a form as $\Omega_\chi h^2\propto m_\chi Y_D\propto m_\chi/\lambda_\Phi$. This result is also insensitive to the choice of $\lambda_S(m_S)$ and $\lambda_{HS}(m_S)$ since the dominant production process is the RIS one. As the $\lambda_\Phi$ decreases, all the contours converge to the same line corresponding to $\lambda_\Phi m_\chi\sim10^{-17}$~GeV. It can also be understood by considering Eq.\eqref{approxYD} in the limit of $\lambda_\Phi\ll\lambda_{H\Phi},\lambda_{\Phi S}$, then
\begin{eqnarray}
Y_D\approx\frac{45M_P\lambda_\Phi}{64\pi^6 m_\phi}\sqrt{\frac{g_\ast}{90}}\int_{m_\phi/T_R}^{\infty}dx_\phi x_\phi^3K_1\left(x_\phi\right)\sqrt{1-\left(\frac{2x_\chi}{x_\phi}\right)^2},
\end{eqnarray}
so $\Omega_\chi h^2\propto \lambda_\Phi m_\chi=$~constant is the asymptotic line for different contours, and the results are also insensitive to the choice of $\lambda_S(m_S)$ and $\lambda_{HS}(m_S)$. However, as the $\lambda_\Phi$ further decreases and reaches some critical value, the dominant production process will become the t,u-channels of $\overline{H}+ H(S+S)\to\phi+\phi$, then the approximated yield given by Eq.\eqref{approxYD} is invalid. In this parameter regions, $\Omega_\chi h^2\propto m_\chi/\lambda_\Phi^2$ for fixing $\lambda_{H\Phi}$ and $\lambda_{\Phi S}$, and the results are sensitive to the chosen $\lambda_S(m_S)$ and $\lambda_{HS}(m_S)$. In conclusion, the relic abundance of DM in this model is sensitive to the details of $H$-$S$ mixing only when $(\lambda_{H\Phi}^2/\lambda_\Phi)$,~or $(\lambda_{\Phi S}^2/\lambda_\Phi)\gtrsim10^{-4}$.

\begin{figure}[!t]
	\centering
	\subfigure[\label{fig8-1}]
	{\includegraphics[width=0.49\textwidth]{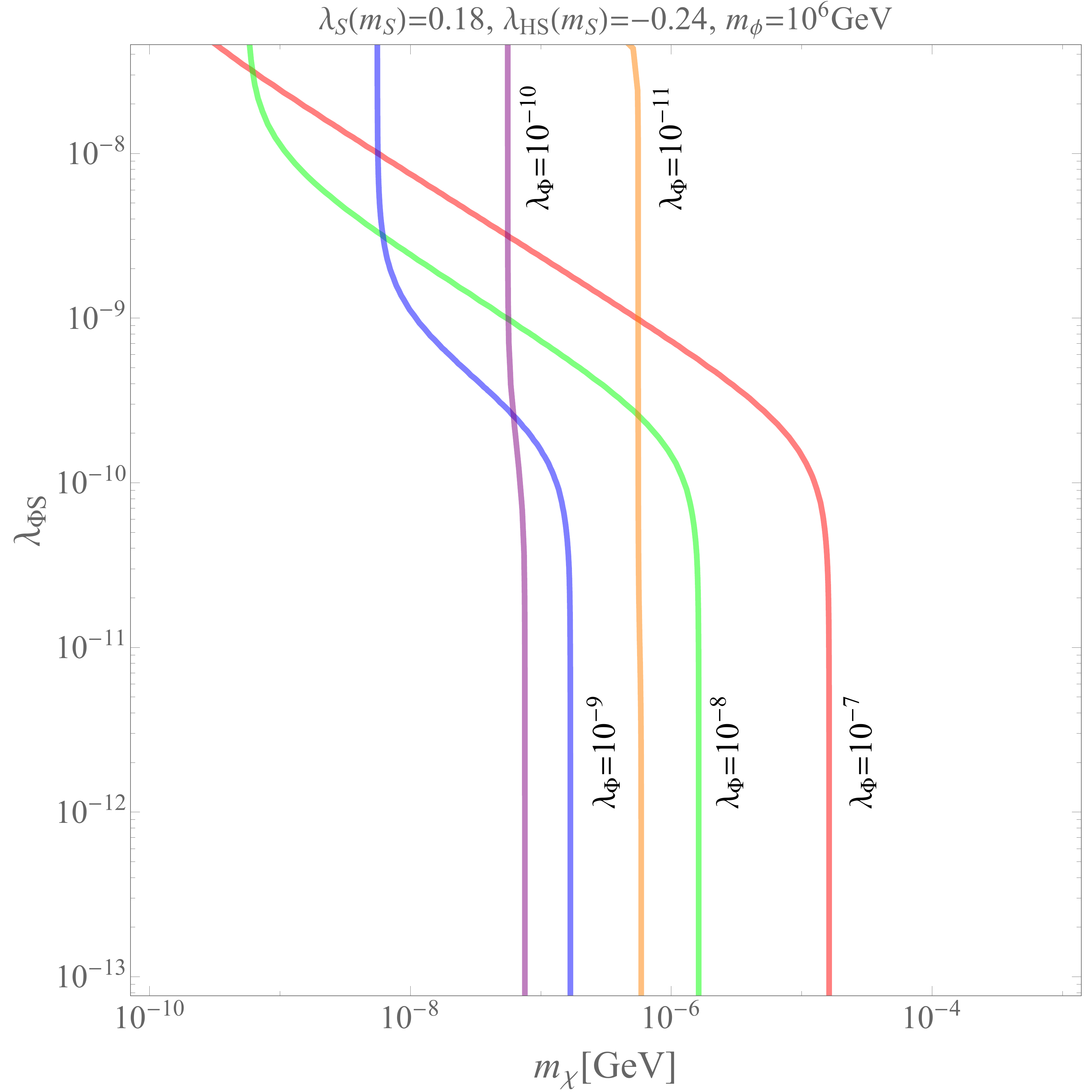}}
	\subfigure[\label{fig8-2}]
	{\includegraphics[width=0.49\textwidth]{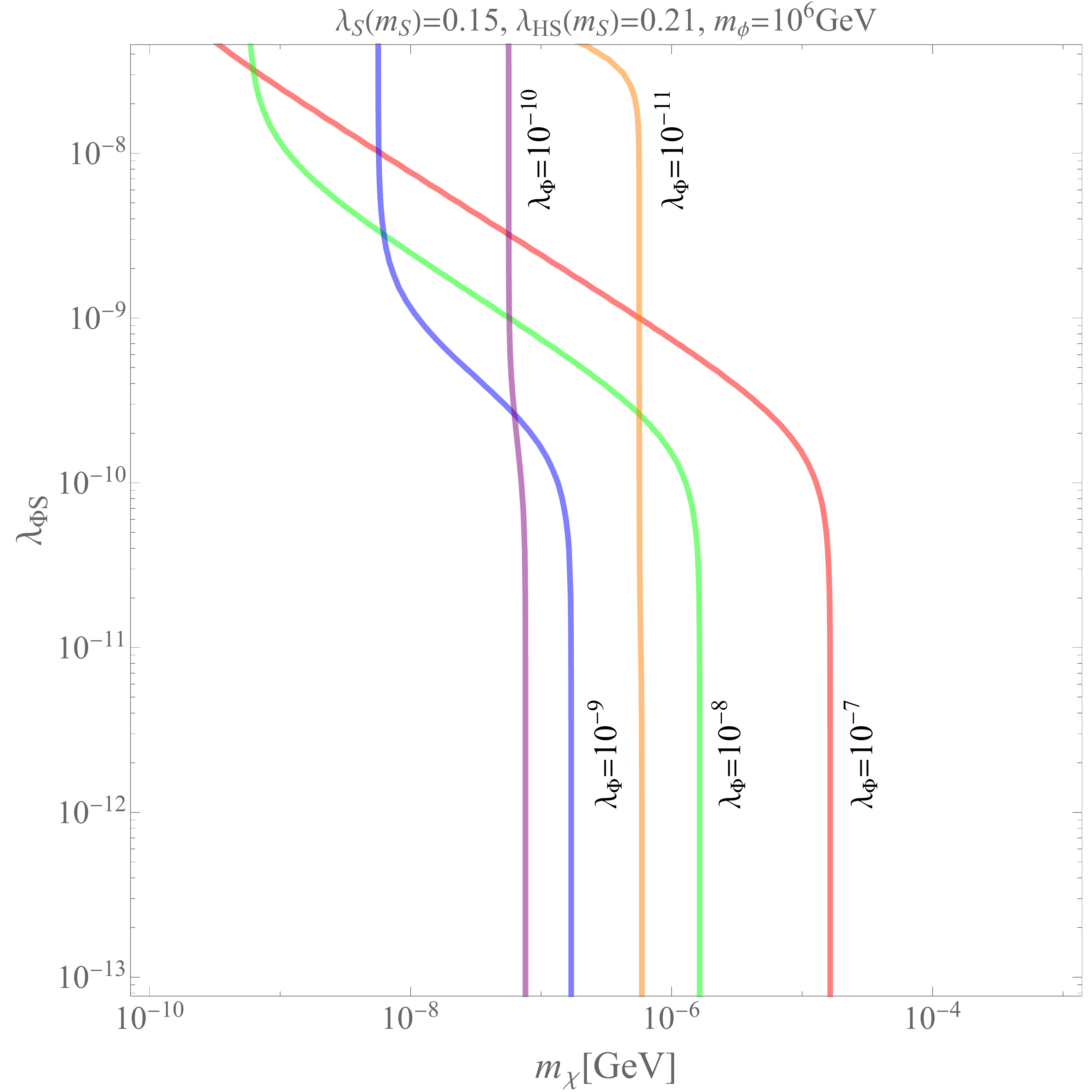}}
	\caption{Contours reproducing the observed relic abundance in the $(m_\chi,\lambda_{\Phi S})$
		plane with $m_\phi = 10^6$GeV, $\lambda_{H \Phi} = 10^{-10}$ and  $\lambda_{\Phi} = 10^{-7}\sim10^{-11}$.
		The red, green, blue, purple, orange lines correspond to  $\lambda_{\Phi} = 10^{-7},10^{-8}, 10^{ -9},10^{ -10}$ and $10^{ -11}$ respectively.
		(Left):$\lambda_S(m_S) = 0.18,\lambda_{HS}(m_S) = -0.24$. (Right): $\lambda_S(m_S) = 0.15,\lambda_{HS}(m_S) = 0.21$ 	}
	\label{fig8}
\end{figure}

\begin{figure}[!t]
	\centering	
	{\includegraphics[width=0.7\textwidth]{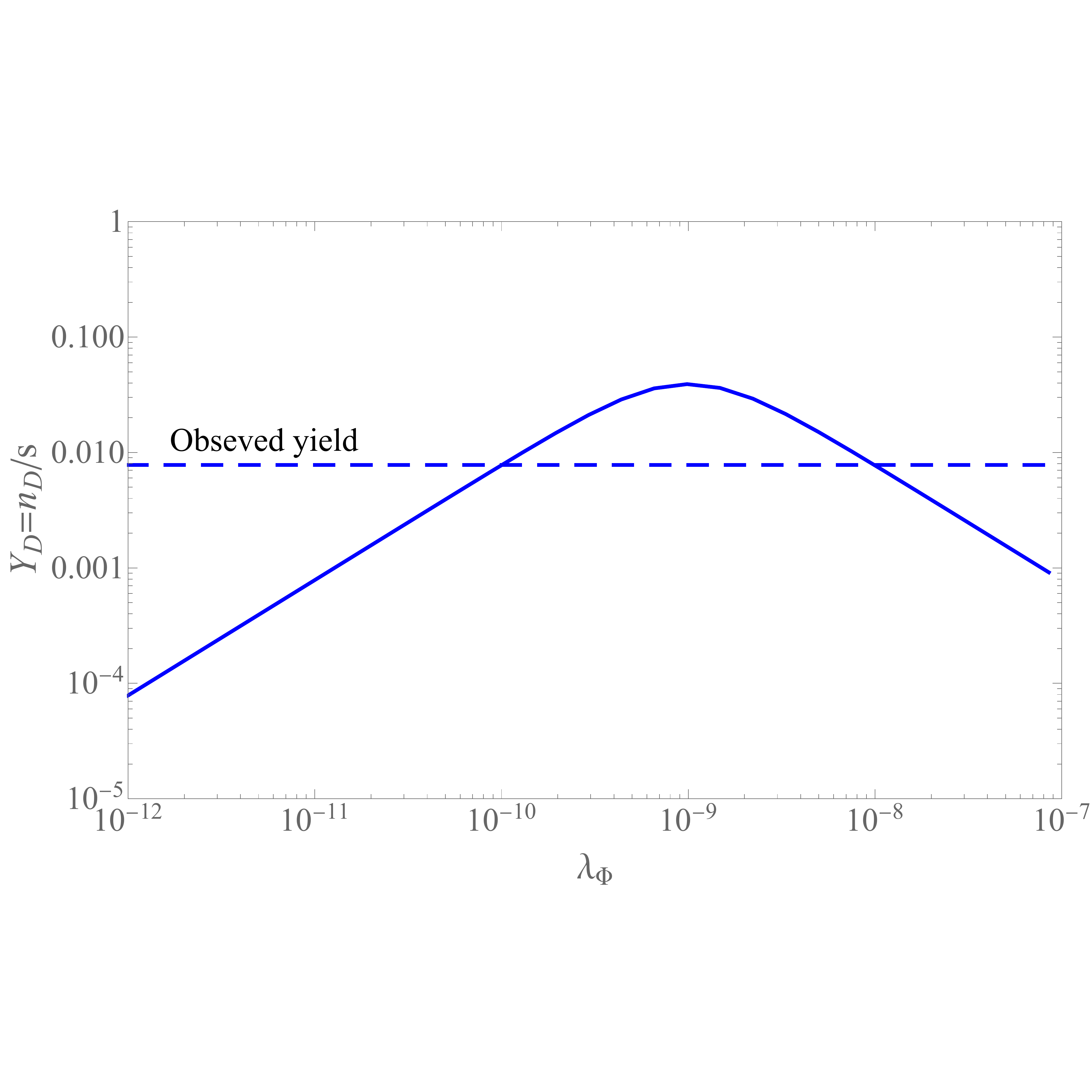}}
	\caption{ The yield of DM evolving with $\lambda_\Phi$ for parameters at the cross-over point on the green and purple line in Fig.~\ref{fig8}.
	}
	\label{fig9}
\end{figure}

In FIG.\ref{fig8}, we show the contours which can reproduce the observed DM relic abundance in the  $m_\chi$ vs. $\lambda_{\Phi S}$ plane.
The parameters are chosen as the same as FIG.\ref{fig7} except that $\lambda_{H \Phi}$ and $\lambda_{ \Phi}$ are set as
\begin{eqnarray}
\lambda_{H\Phi} = 10^{-10}, \quad \lambda_\Phi = 10^{-7} \sim 10^{ -11},
\end{eqnarray}
and $\lambda_{\Phi S}$ is chosen as a free parameter.
Note that there are some cross-over points for some lines, which corresponds to the same set of parameters but different $\lambda_{ \Phi}$ of the model leading to the observed relic abundance. In order to understand the behavior of $Y_D$ along the $\lambda_\Phi$ direction, we plot $Y_D$ as a function of $\lambda_\Phi$ in FIG.~\ref{fig9} with fixing
\begin{eqnarray}
m_\chi \approx  5.65\times10^{-18}~\text{GeV},\quad \lambda_{\Phi S} \approx 9.82\times10^{-10},
\end{eqnarray}
which corresponds to the cross-over point in Fig.~\ref{fig8} of the green and purple contours. We can see that $Y_D$ is not a monotonic function of $\lambda_\Phi$. Therefore, there can be two points on the line, $\lambda_\Phi = 10^{-10} $ and $ 10^{ -8} $, which can result in the observed relic abundance.

\section{Summary}\label{se.summary}

A pseudo-Nambu-Goldstone dark matter model extended with a real scalar is studied in this work. A $Z_2$ symmetry is introduced for simplifying the potential terms. The complex scalar field $\Phi$ is assumed to be feebly coupled with the SM sector and the real scalar $S$, so both the radial and phase components of $\Phi$ are FIMPs and they can be produced via the freeze-in mechanism after reheating. The real scalar field $S$ is introduced for stabilizing the electroweak vacuum, so the constraints from vacuum stability and perturbativity of couplings up to the Planck scale can be used as theoretical constraints of our model. These constraints force the quartic couplings $\lambda_S(m_S)$ and $\lambda_{HS}(m_S)$ at the mass scale of $m_S=10$~TeV to lie in a restricted region that $0<\lambda_S(m_S)\lesssim0.25$ and $-0.39\lesssim\lambda_{HS}(m_S)\lesssim0.34$.

In this work, we focus on the case of IR freeze-in production, which means the reheating temperature $T_R$ is much larger than the mass of $\Phi$ and the final yield of DM is insensitive to $T_R$. In order to include the effect of running couplings, we use an approximation that all the couplings are fixed to their values at $\mu\sim m_\phi$ when we solve the Boltzmann equations numerically. We find that if $\lambda_{H\Phi},\lambda_{\Phi S}\lesssim \lambda_{\Phi}$, the dominant production processes are $\overline{H}+H\to\phi\to\chi+\chi$ and $S+S\to\phi\to\chi+\chi$ with a real intermediated $\phi$ state. In this situation, the yield of DM is insensitive to the value of couplings $\lambda_S(m_\phi)$ and $\lambda_{HS}(m_\phi)$ in the scale of freeze-in production. We also considered the situation of $\lambda_{H\Phi},\lambda_{\Phi S}\gg \lambda_{\Phi}$ which are less discussed in the previous study~\cite{Abe20}. We find that the dominant DM production processes become $\overline{H}+H\to \phi+\phi$ and $S+S\to \phi+\phi$ via the t- and u-channels.
We also find that the yield of DM is sensitive to the value of $\lambda_S(\mu)$ and $\lambda_{HS}(\mu)$ in this situation.
In all these cases, the observed relic abundance of DM can be reproduced with proper parameter sets.

\begin{acknowledgments}
This work is supported by the National Natural Science Foundation of China (NSFC) under Grants No. 12275367, No. 11905300 and No. 11875327, the Fundamental Research Funds for the Central Universities, the Natural Science Foundation of Guangdong Province, and the Sun Yat-Sen University Science Foundation.
\end{acknowledgments}

%%%%%%%%%%%
\appendix
\section{The effect of including t- and u-channels of the process $\overline{H} +H \to \phi +\phi$ in the SM+pNGB Model}\label{app1}

In this appendix, we compare the freeze-in production results of pNGB model with and without t- and u-channels in the $\overline{H} +H \to \phi+ \phi$ process.

\begin{figure}[!t]
	\centering	
	{\includegraphics[width=0.8\textwidth]{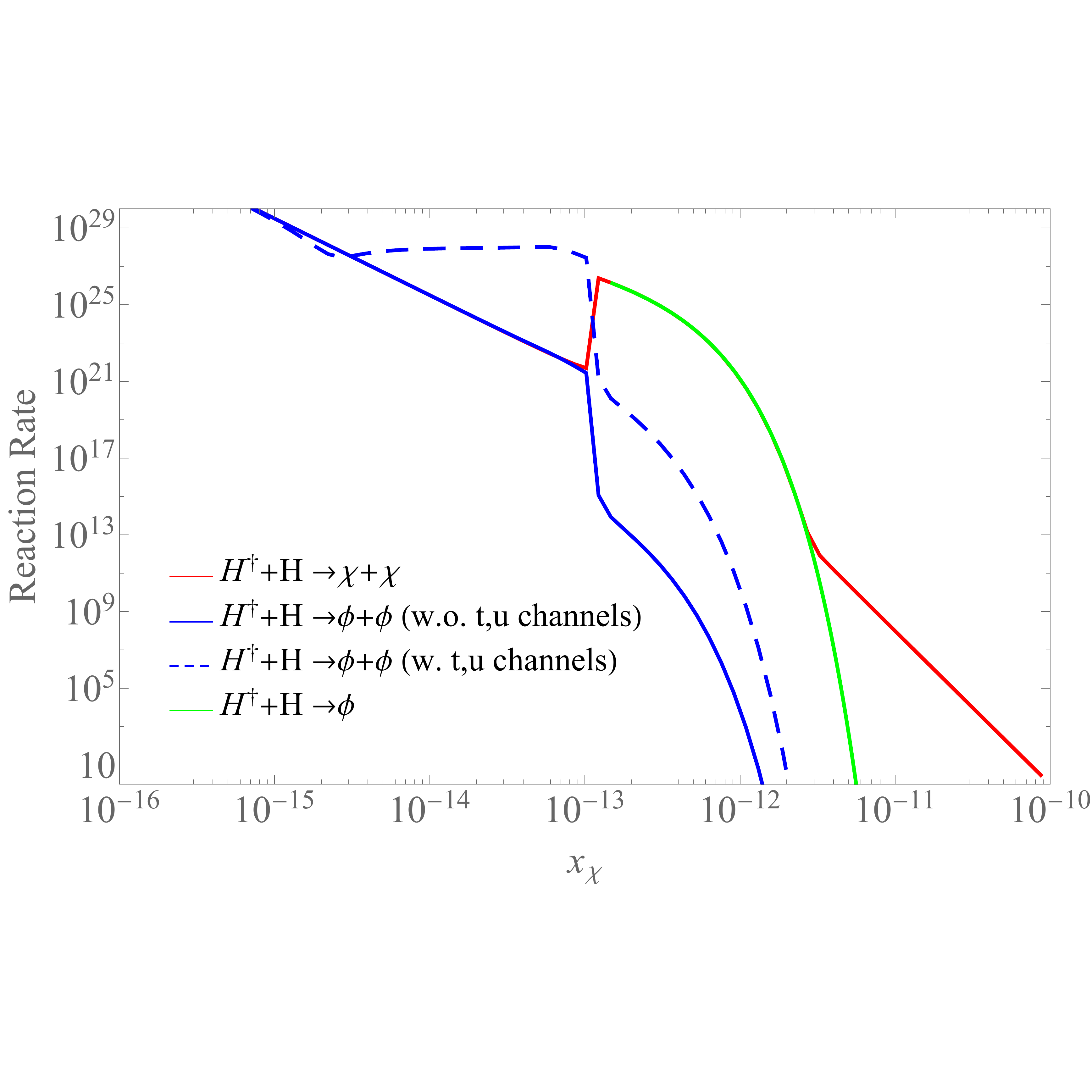}}
	\caption{Evolution of the reaction rates for the different processes of the SM+pNGB model.}
	\label{fig1}
\end{figure}

The reaction rate changing with $x_\chi = m_\chi/T$ is given in Fig.~\ref{fig1}. The benchmark parameters are the same as those in Ref.\cite{Abe20}, which are
\begin{eqnarray}
\lambda_\Phi = 7 \times 10^{-11},\quad \lambda_{H\Phi} = 10^{-7},\quad m_\chi = 1~\text{MeV},\quad m_\phi = 10^{10}~\text{GeV}.
\end{eqnarray}
The green and red lines denote the reaction rates for processes $\overline{H} +H \to \chi+\chi$ and $\overline{H} +H \to \phi$ respectively. The solid and dashed blue lines denote the reaction rates for process $\overline{H} +H \to \phi+\phi$ without and with the t- and u-channels, respectively.  We can see that the rate including these channels is significantly enhanced in the range $x_\chi \gtrsim 10^{-15} $  comparing to the rate ignoring $t,u$ channels. It means that the t- and u-channels are not always negligible. According to Eq.(\ref{ampHHphiphi}), we find that these channels should be taken into account when $\lambda_{H \Phi}^2 \gg \lambda_\phi^2$

\begin{figure}[!t]
	\centering	
	{\includegraphics[width=0.8\textwidth]{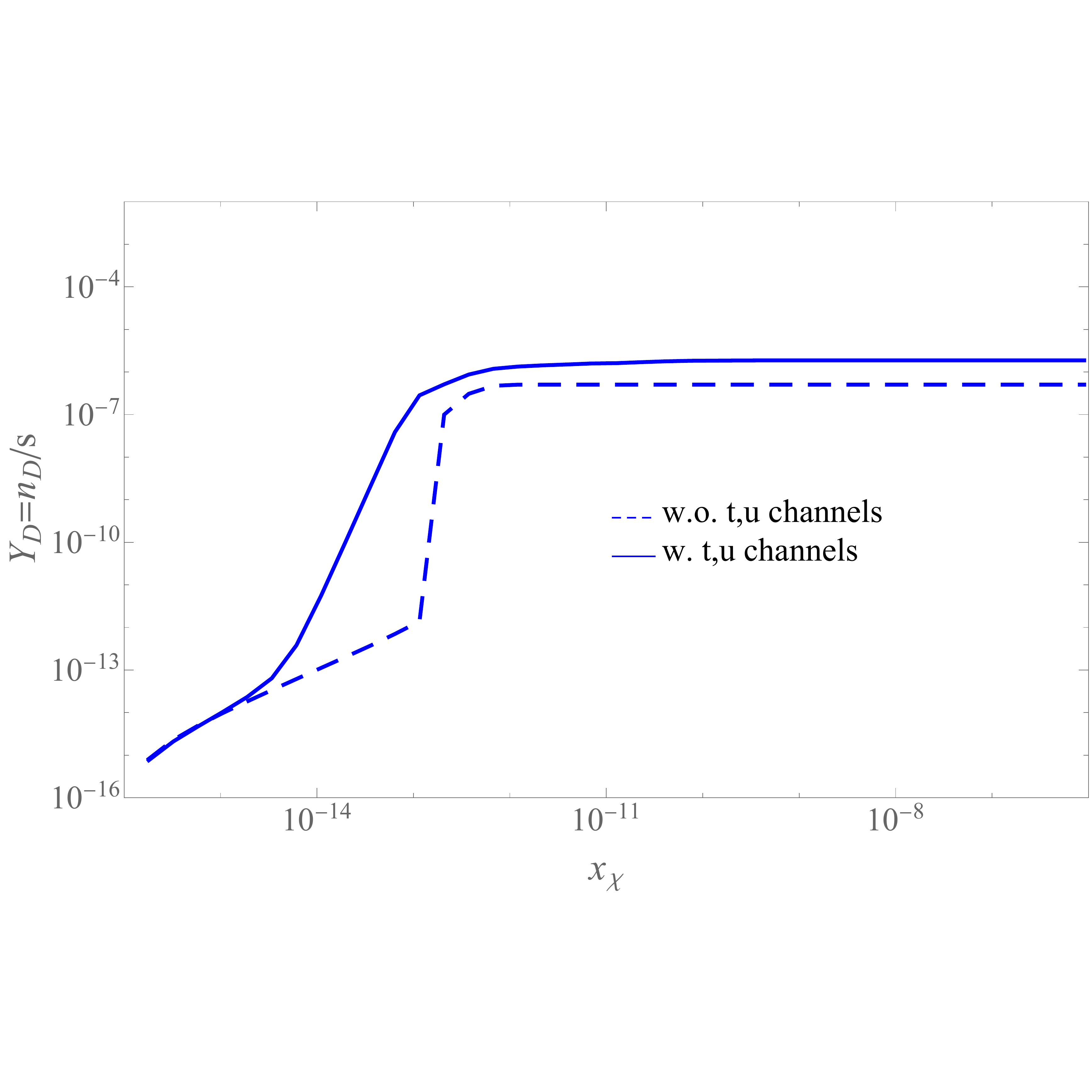}}
	\caption{Comparison of the dark matter yield including and not including the  t- and u-channels of the process $\overline{H}+H \to \phi+\phi$.}
	\label{fig2}
\end{figure}
%%%%%%%
Since the t- and u-channels of $\overline{H} +H \to \phi+\phi$ might become the dominant parts of the DM production in the range around  $10^{-15}<x_\chi<10^{-13}$, the evolution of the yield of DM can also be significantly changed. In Fig.~\ref{fig2}, we show the comparison of the yields corresponding to the computation with and without the t- and u-channels. We find that including these channels can lead to an enhancement with approximately an order of magnitude comparing to the results of ignoring them.

\begin{figure}[!t]
	\centering	
	{\includegraphics[width=0.7\textwidth]{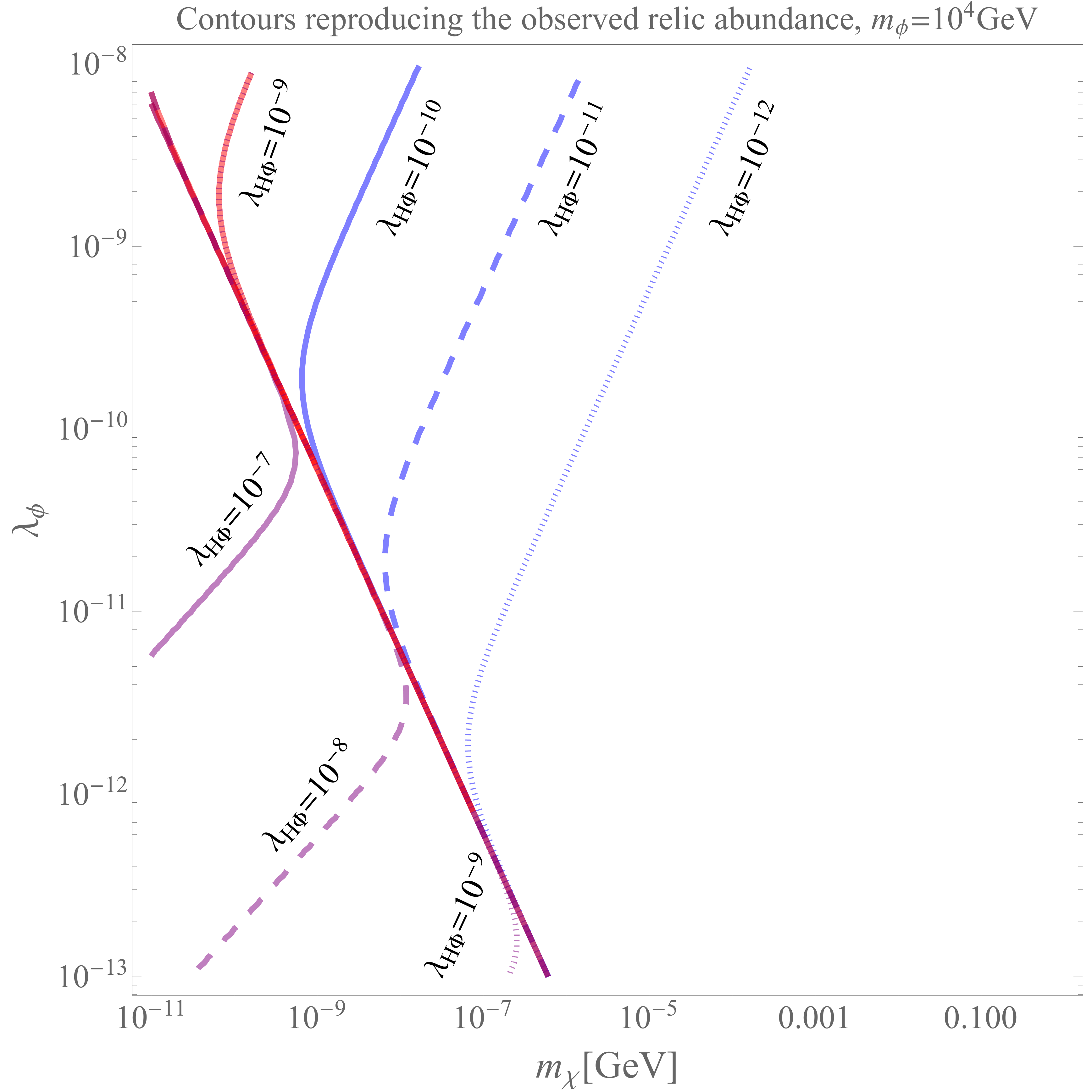}}
	\caption{Contours reproducing the observed relic abundance in the $(m_\chi,\lambda_\phi)$
		plane with $m_\phi = 10^4$~GeV for $\lambda_{H \Phi} = 10^{-12} - 10^{-7}$. 		
		The blue solid, dashed and dotted line corresponds to results for  $\lambda_{H
			\Phi} = 10^{-10}, 10^{-11}$ and $10^{-12}$ respectively, including t- and u-channels of $\overline{H}+H \to \phi+\phi$.
		The purple solid, dashed and dotted line corresponds to results for $\lambda_{H \Phi} = 10^{-7},~10^{-8}$  and $10^{-9}$ respectively, including t- and u-channels of $\overline{H}+H \to \phi+\phi$.
		The red lines corresponds to the results for $\lambda_{H \Phi} = 10^{-9}, 10^{-8}$  and $10^{-7}$  without t- and u-channels.
	}
	\label{fig3}
\end{figure}
Fig.~\ref{fig3} shows the contours in the $m_\chi$ vs. $\lambda_\phi$ plane correspond to the parameter sets which can reproduce the relic abundance of DM observed by the PLANCK Collaboration \cite{Planck:2018vyg}.
We only consider the IR freeze-in case in which the reheating temperature is much higher than the mediator mass ($T_R \gg m_\phi$). The blue lines indicate the contours favored by the relic abundance of DM for $\lambda_{H \Phi}=10^{-12},~10^{-11},~10^{-10}$. These results are almost the same as the ones in Ref.\cite{Abe20} since $\lambda_{H\Phi}^2\lesssim\lambda_\Phi^2$ in these cases. Therefore, the effects of including t- and u-channels are negligible for $\lambda_{H \Phi}=10^{-12}-10^{-10}$. However, for larger $\lambda_{H\Phi}$, such as $\lambda_{H\Phi}=10^{-9} - 10^{-7}$, results with t- and u-channels (denoted by purple lines), can deviate a lot from the results without  t- and u-channels (denoted by the red lines).

\section{$\beta$-functions of the standard model} \label{se.beta}
The evolutions of SM couplings are determined by the renormalization group equations (RGE).
The $\beta$ functions for SM up to two-loop level are listed below~\cite{Cheng:1973nv,Machacek:1983tz,Machacek:1983fi,Machacek:1984zw,Arason:1991ic}
\begin{eqnarray}\label{betagi}
\beta_{g_{i}}
&=&
\frac{1}{(4 \pi)^{2}} g_{i}^{3} b_{i}+\frac{1}{(4 \pi)^{4}} g_{i}^{3}\left[\sum_{j=1}^{3} c_{i j} g_{j}^{2}-d_{i} y_{t}^{2}\right],
\end{eqnarray}
with

\begin{eqnarray}
g_{i}=\left\{g^{\prime},
g, g_{s}\right\},
b=\left(\frac{41}{6},-\frac{19}{6},-7\right),
c=\left(
\begin{array}{ccc}
\frac{199}{18} & \frac{9}{2} & \frac{44}{3} \nonumber \\
\frac{3}{2} & \frac{35}{6} & 12 \nonumber \\
\frac{11}{6} &\frac{9}{2} & -26
\end{array}
\right),
d=\left(\frac{17}{6}, \frac{3}{2}, 2\right).
\end{eqnarray}
\\
and
\begin{eqnarray}
\beta_{\lambda_H}
&=&
\frac{1}{(4 \pi)^{2}}
\left[24 \lambda_H^{2}-6 y_{t}^{4}+\frac{3}{8}\left(2 g^{4}+\left(g^{2}+g^{\prime 2}\right)^{2}\right)+\left(-9 g^{2}-3 g^{\prime 2}+12 y_{t}^{2}\right) \lambda_H
\right]
\nonumber\\
&\quad+&\frac{1}{(4 \pi)^{4}}\left[\frac{1}{48}\left(915 g^{6}-289 g^{4} g^{\prime 2}-559 g^{2} g^{\prime 4}-379 g^{\prime 6}\right)+30 y_{t}^{6}-y_{t}^{4}\left(\frac{8 g^{\prime 2}}{3}+32 g_{s}^{2}+3 \lambda_H\right)\right.
\nonumber\\
&\quad+&\lambda\left(-\frac{73}{8} g^{4}+\frac{39}{4} g^{2} g^{\prime 2}+\frac{629}{24} g^{\prime 4}+108 g^{2} \lambda+36 g^{\prime 2} \lambda_H-312 \lambda_H^{2}\right)
\nonumber\\
&\quad+&
\left.y_{t}^{2}
\left(-\frac{9}{4} g^{4}+\frac{21}{2} g^{2} g^{\prime 2}-\frac{19}{4}
g^{\prime 4}+\lambda_H\left(\frac{45}{2} g^{2}+\frac{85}{6} g^{\prime 2}+80 g_{s}^{2}-144 \lambda_H\right)
\right)
\right],
\label{betalam}
\\
\beta_{y_{t}}
&=&
\frac{y_{t}}{(4 \pi)^{2}}\left[\frac{9}{2} y_{t}^{2}-\frac{9}{4} g^{2}-\frac{17}{12} g^{\prime 2}-8 g_{s}^{2}\right]+\frac{y_{t}}{(4 \pi)^{4}}\left[-\frac{23}{4} g^{4}-\frac{3}{4} g^{2} g^{\prime 2}+\frac{1187}{216} g^{\prime 4}+9 g^{2} g_{s}^{2}\right.
\nonumber\\
&&\left.+
\frac{19}{9} g^{\prime 2} g_{s}^{2}-108 g_{s}^{4}+\left(\frac{225}{16} g^{2}+\frac{131}{16} g^{\prime 2}+36 g_{s}^{2}\right) y_{t}^{2}+6\left(-2 y_{t}^{4}-2 y_{t}^{2} \lambda_H+\lambda_H^{2}\right)\right] .
\label{betayt}
\end{eqnarray}
Some mass values used in the calculations  are
\begin{eqnarray}
m_{H}=125~\mathrm{GeV}, m_{t}=173~\mathrm{GeV}, m_{Z}=91.188~\mathrm{GeV}, m_{W}=80.2~\mathrm{GeV}, v=246~\mathrm{GeV}.
\end{eqnarray}
The initial values for running couplings $g_i$ are
\begin{eqnarray}
&&\alpha_{s}\left(m_{Z}\right)=\frac{g_{s}^2\left(m_{Z}\right)}{4 \pi}=0.1184,
\nonumber \\
&&\alpha\left(m_{Z}\right)= \frac{g^2\left(m_{Z}\right) s^2_{w}\left(m_{Z}\right)}{4 \pi}=\frac{1}{127.926},
\nonumber \\
&&s_{w}^{2}=\sin ^{2} \theta_{w}\left(m_{Z}\right)=0.2312~.
\end{eqnarray}

The running of $\lambda_H$ and $y_t$ needs the matching condition from Ref.\cite{Hambye:1996wb}.
For $\lambda_H$,
\begin{eqnarray}
\overline{\lambda_H}\left(\mu_{0}\right) &=&
\frac{m_{H}^{2}}{2 v^{2}}\left[1+\delta_{H}\left(\mu_{0}\right)\right]~,
\\
\delta_{H}\left(\mu_{0}\right) &=&\frac{2 v^{2}}{m_{H}^{2}} \frac{1}{32 \pi^{2} v^{4}}\left[h_{0}\left(\mu_{0}\right)+m_{H}^{2} h_{1}\left(\mu_{0}\right)+m_{H}^{4} h_{2}\left(\mu_{0}\right)\right]~,
\\
h_{0}\left(\mu_{0}\right) &=&-24 m_{t}^{4} \ln \frac{\mu_{0}^{2}}{m_{t}^{2}}+6 m_{Z}^{4} \ln \frac{\mu_{0}^{2}}{m_{Z}^{2}}+12 m_{W}^{4} \ln \frac{\mu_{0}^{2}}{m_{W}^{2}}+c_{0} ~,
\\
h_{1}\left(\mu_{0}\right) &=&12 m_{t}^{2} \ln \frac{\mu_{0}^{2}}{m_{t}^{2}}-6 m_{Z}^{2} \ln \frac{\mu_{0}^{2}}{m_{Z}^{2}}-12 m_{W}^{2} \ln \frac{\mu_{0}^{2}}{m_{W}^{2}}+c_{1} ~,
\\
h_{2}\left(\mu_{0}\right) &=&\frac{9}{2} \ln \frac{\mu_{0}^{2}}{m_{H}^{2}}+\frac{1}{2} \ln \frac{\mu_{0}^{2}}{m_{Z}^{2}}+\ln \frac{\mu_{0}^{2}}{m_{W}^{2}}+c_{3}~.
\end{eqnarray}
For $y_t$,
\begin{eqnarray}
y_{t}\left(\mu_{0}\right)&=&\frac{\sqrt{2} m_{t}}{v}\left[1+\delta_{t}\left(\mu_{0}\right)\right] ~,
\nonumber \\
\delta_{t}\left(\mu_{0}\right)&=&\left(-\frac{4 \alpha_{s}}{4 \pi}-\frac{4}{3} \frac{\alpha}{4 \pi}+\frac{9}{4} \frac{m_{t}^{2}}{16 \pi^{2} v^{2}}\right) \ln \frac{\mu_{0}^{2}}{m_{t}^{2}}+c_{t}~,
\end{eqnarray}
where $-0.052<c_{t}<-0.042$.

\bibliographystyle{utphys}
\bibliography{ref}

\end{document}